%% file: main.tex
\pdfoutput=1
\documentclass[fleqn,usenatbib]{mnras}

\usepackage{mathptmx}

\usepackage[T1]{fontenc}
\usepackage{ae,aecompl}

\usepackage{graphicx}	
\usepackage{amsmath}	
\usepackage{amssymb}	
\usepackage{multirow}
\usepackage{hyperref}

\newcommand{\app}[1]{Appendix~\ref{sec:#1}}

\newcommand{\fig}[1]{Figure~\ref{fig:#1}}
\renewcommand{\sec}[1]{Section~\ref{sec:#1}}
\newcommand{\tab}[1]{Table~\ref{tab:#1}}

\title[Learning Star Formation Histories]{Learning the Relationship between Galaxies Spectra and their Star Formation Histories using Convolutional Neural Networks and Cosmological Simulations}

\author[C. C. Lovell et al.]{Christopher C. Lovell,$^{1,2}$\thanks{E-mail: c.lovell@herts.ac.uk}
Viviana Acquaviva,$^{3}$
Peter A. Thomas,$^{1}$
\newauthor
Kartheik G. Iyer,$^{4}$
Eric Gawiser,$^{4,5}$
Stephen M. Wilkins$^{1}$
\\
$^{1}$Astronomy Centre, Department of Physics and Astronomy, University of Sussex, Brighton, BN1 9QH, UK\\
$^{2}$Centre for Astrophysics Research, School of Physics, Astronomy \& Mathematics, University of Hertfordshire,\\Hatfield, AL10 9AB, UK\\
$^{3}$Department of Physics, New York City College of Technology, Brooklyn, NY 11201, USA\\
$^{4}$Department of Physics and Astronomy, Rutgers, The State University of New Jersey, 136 Frelinghuysen Road,\\Piscataway, NJ 08854-8019 USA\\
$^{5}$Center for Computational Astrophysics, Flatiron Institute, 162 5th Ave, New York, NY 10010, USA
}

\date{Accepted 2019 October 04 . Received 2019 September 20; in original form 2019 March 25}

\pubyear{2019}

\begin{document}
\label{firstpage}
\pagerange{\pageref{firstpage}--\pageref{lastpage}}
\maketitle

\begin{abstract}
We present a new method for inferring galaxy star formation histories (SFH) using machine learning methods coupled with two cosmological hydrodynamic simulations.
We train Convolutional Neural Networks to learn the relationship between synthetic galaxy spectra and high resolution SFHs from the \textsc{EAGLE} and Illustris models.
To evaluate our SFH reconstruction we use Symmetric Mean Absolute Percentage Error (SMAPE), which acts as a true percentage error in the low-error regime.
On dust-attenuated spectra we achieve high test accuracy (median SMAPE = 10.5\%).
Including the effects of simulated observational noise increases the error (12.5\%), however this is alleviated by including multiple realisations of the noise, which increases the training set size and reduces overfitting (10.9\%).
We also make estimates for the observational and modelling errors.
To further evaluate the generalisation properties we apply models trained on one simulation to spectra from the other, which leads to only a small increase in the error (median SMAPE $\sim 15\%$).
We apply each trained model to SDSS DR7 spectra, and find smoother histories than in the VESPA catalogue.
This new approach complements the results of existing SED fitting techniques, providing star formation histories directly motivated by the results of the latest cosmological simulations.
\end{abstract}




\input{intro.tex}
\input{methods.tex}

\input{methods_ml.tex}
\input{methods_sims.tex}
\input{methods_spectra.tex}
\input{results.tex}
\input{testing.tex}
\input{errors.tex}
\input{methods_obs.tex}
\input{obs_comparison.tex}
\input{conc.tex}

\section*{Acknowledgements}
The authors wish to thank the anonymous referee for providing useful comments and suggestions that improved this manuscript.
We would also like to thank Rita Tojeiro for help understanding \textsc{Vespa}.
We wish to acknowledge the use of the following open source software packages not mentioned directly in the text: Scipy \citep{scipy} and Astropy \citep{astropy}.
CCL (ORCID 0000-0001-7964-5933) acknowledges the support of a PhD studentship from the UK Science and Technology Facilities Council (STFC).
VA, KI \& EG acknowledge that support for program number HST-AR-14564.001-A and GO-12060 was provided by NASA through a grant from the Space Telescope Science Institute, which is operated by the Association of Universities for Research in Astronomy, Incorporated, under NASA contract NAS5-26555.
The Flatiron Institute is supported by the Simons Foundation.



\bibliographystyle{mnras}
\bibliography{SFHs,extra,Classics}



\appendix
\input{appendix.tex}


\bsp	
\label{lastpage}
\end{document}

%% file: intro.tex
\section{Introduction}
\label{sec:intro}

A galaxy's integrated Spectral Energy Distribution (SED) contains information about countless physical properties, such as the stellar population age, mass, dust content, redshift, metallicity and star formation history (SFH). Different physical processes leave their imprint in different parts of the spectrum; the wider and more finely sampled the wavelength coverage, the more robust the interpretation of the various features of the SED is in terms of galaxy properties.
One fundamental tool to determine the physical properties of a galaxy starting from photometric and/or spectroscopic observations is SED fitting, the procedure of iteratively comparing models to the observed galaxy SEDs (e.g. \citealt{walcher_fitting_2011, conroy_modeling_2013}). Since the physical properties of the models are known, those of the data can be derived by maximizing the resemblance between data and models. The success and reliability of this method depends on the quality of the available template spectra, and the robustness of the fitting algorithm.

The field of SED fitting has seen enormous progress in the last decade \citep{conroy_modeling_2013}.
Methods such as Markov Chain Monte Carlo have been used to efficiently explore the degeneracies associated with the large parameter space (e.g. \citealt{sajina_1-1000m_2006, acquaviva_spectral_2011, pirzkal_link_2012, acquaviva_sed_2012, leja_deriving_2017}).
However, one issue that has consistently emerged from these efforts is the difficulty of characterizing and constraining the star formation histories of galaxies.
The spectral signatures of multiple non-coeval generations of stars can be mimicked by other physical effects, such as varying stellar metallicity, and older stellar populations with high mass-to-light ratios are easily hidden in observed spectra, an effect sometimes referred to as ``outshining" \citep{maraston_star_2010}.
It would be helpful, in Bayesian parameter estimation, to use priors to guide our exploration of very large and degenerate parameter spaces, but these are not readily available.

A wrongly reconstructed star formation history introduces significant biases in many parameters that are usually estimated through Spectral Energy Distribution fitting, such as stellar masses, stellar age indicators, dust content, and redshift (e.g. \citealt{mobasher_critical_2015, pacifici_importance_2014, iyer_reconstruction_2017, leja_deriving_2017}).
\cite{acquaviva_simultaneous_2015} evaluated the impact of different sources of non-algorithmic systematics on the recovered SED fitting parameters and concluded that a wrong star formation history is the most detrimental.
Similarly, \cite{iyer_reconstruction_2017} found that fitting the SFH using single stellar populations and simple functional forms (e.g. exponentially declining or constant models) leads to a bias of up to $70\%$ in the recovered total stellar mass.
\cite{carnall_how_2019} further demonstrated that simple parametric star formation histories impose strong priors on implied physical parameters.
These introduce strong correlated biases that are propagated through pipelines of results and used to infer key distribution functions and relations, such as the stellar mass function and the cosmic star formation rate density \citep{ciesla_sfr-m_2017,leja_how_2019}, critical for answering crucial questions in the study of galaxy formation and evolution.

One possible approach to solving this problem has been to introduce new parametrisations for the SFH that are less subject to the outshining bias \citep{behroozi_average_2013, simha_parametrising_2014}, or to develop parameter-free descriptions of the SFH \citep{tojeiro_recovering_2007, iyer_reconstruction_2017, iyer_nonparametric_2019, leja_how_2019}.
Here we propose an alternative approach, using supervised machine learning algorithms to `learn' the relationship between the SFH and the SEDs of galaxies.
In contrast with SED fitting, where the SFH is built from some ensemble of simple stellar populations to maximise the resemblance in SED space, machine learning directly learns the relationship between the spectra and the entire SFH.
We expect that this method will carry systematic uncertainties that are independent of those from SED fitting, so that our results will complement and strengthen the results of these approaches.
Another strength of a machine learning-based approach is that the algorithm learns from the population ensemble, learning not only the correspondence between individual spectra and star formation histories, but also which star formation histories are common and which are unlikely, something that would be analogous in Bayesian parameter estimation to learning the SFH prior.

A number of recent studies have explored the effect of priors on derived SFHs in SED fitting approaches.
\cite{carnall_how_2019} showed that parametric approaches implicitly impose a strong prior on the SFH that can lead to unrealistically tight posterior constraints on the SFR, and \cite{leja_how_2019} showed that even non-parametric fits are sensitive to the prior SFH distribution, particularly where the data are poor.
\cite{pacifici_rise_2013} proposed using SFHs from a semi-analytic model to generate a library of SEDs to be used in an SED fitting algorithm, and found that these simulation-motivated templates prefer symmetric or rising SFHs at intermediate redshifts ($0.2 < z < 1.4$), compared to the exponentially declining forms predicted using simple stellar populations.
Finally, \cite{wilkins_single-colour_2013} show that using simulation-motivated enrichment and star-formation histories leads to more accurate stellar mass estimates from colour information only.
These studies highlight the importance of the explicitly or implicitly assumed prior distribution of SFHs.

Machine learning methods are becoming an increasingly popular tool for Astronomers \citep{ball_data_2010,baron_machine_2019}. This is particularly the case where there is abundant low quality data for which expensive, higher quality data can be obtained and used for supervised training.
However, a supervised machine learning algorithm is only as good as its learning sample.
The main challenges to applying these techniques to measure properties such as star formation histories have been the following: assembling a sample of galaxies for which the ``true'' star formation history is known; and making sure that properties of the ensemble (the distribution of properties and their relationship to one another) are a fair snapshot of the real Universe.
However, there has been significant recent progress from multiple independent teams on high-resolution cosmological models of galaxy evolution, which has for the first time provided the potential to test this technique \citep[e.g.][]{simet_comparison_2019}.

Hydrodynamic cosmological simulations in particular are able to resolve stellar populations, producing realistic, high resolution SFHs by taking into account a number of effects, such as environmental interactions, mergers, and stellar and AGN feedback \citep{somerville_physical_2015}.
EAGLE \citep{schaye_eagle_2014} and Illustris \citep{genel_introducing_2014} are two recent hydrodynamic simulations that reproduce a number of key galaxy distribution functions.
Both are necessarily tuned to a small number of observational constraints due to their limited resolution, which requires subgrid models to model physical processes below the simulation scale.
Despite this, a number of observables not included in the tuning are simultaneously reproduced. Of interest for this study are the distributions of colours and photometric magnitudes, which are well reproduced in both models \citep{trayford_colours_2015, torrey_synthetic_2015}.
The recent convergence of such detailed models with the observations, and within sufficiently large simulated volumes, has finally enabled them to be used as training sets for machine learning models.

In \sec{method} we describe the method in detail, including an overview of the machine learning techniques (\ref{sec:ml_methods}), the simulations used (\ref{sec:sims}) and our method for generating synthetic spectra with \textsc{spectacle}, a stand-alone python module for generating spectra from cosmological simulations (\ref{sec:spectra}).\footnote{\href{https://github.com/christopherlovell/spectacle}{https://github.com/christopherlovell/spectacle}}
Our results when trained and tested on the simulations are presented in \sec{results}. \sec{errors} details our modelling of the uncertainty contribution from the observational and modelling sources. We then apply our trained models to SDSS observations: \sec{obs_methods} details the selection of our observational sample, \sec{obs_vespa} describes the VESPA SFH catalogue, and \sec{obs_pred} presents our predictions.
Finally, in \sec{disc} we discuss our results and avenues for future research, then summarise our conclusions in \sec{conc}.
We make all of our code for downloading the simulation and observational data, as well as training the CNNs, available online in the form of Jupyter notebooks.\footnote{\href{https://github.com/christopherlovell/learning\_sfhs}{https://github.com/christopherlovell/learning\_sfhs}}
Throughout we assume a Planck 2013 cosmology with the following parameters:
$\Omega_{m} = 0.30$, $\Omega_{\Lambda} = 0.69$, $\Omega_{b} = 0.048$, $h = 0.68$, $\sigma_{8} = 0.83$ and $n_{s} = 0.96$.

%% file: methods.tex
\section{Methodology}
\label{sec:method}

Supervised machine learning methods use training data to learn the relationship between input \textit{features} and output \textit{predictors}. The trained model can then be used to predict values for unseen data.
Our features in this work are galaxy SEDs, and our predictors are SFHs.
We describe the SFHs as a piece-wise constant curve in bins logarithmically spaced in look-back time:
\begin{align}
0 < \; &t \,/\, \mathrm{Myr} \; < 32 \\\nonumber
32 < \; &t \,/\, \mathrm{Myr} \; < 68 \\\nonumber
68 < \; &t \,/\, \mathrm{Myr} \; < 147 \\\nonumber
147 < \; &t \,/\, \mathrm{Myr} \; < 316 \\\nonumber
316 < \; &t \,/\, \mathrm{Myr} \; < 681 \\\nonumber
0.681 < \; &t \,/\, \mathrm{Gyr} \; < 1.47 \\\nonumber
1.47 < \; &t \,/\, \mathrm{Gyr} \; < 3.16 \\\nonumber
3.16 < \; &t \,/\, \mathrm{Gyr} \; < 12.46 \;\;,\nonumber
\end{align}
where $t$ is the lookback time from $z = 0.1$.
This choice ensures that the epochs of recent star formation, which leave more significant imprints on the spectrum, are sampled more finely, while older stellar populations that evolve more slowly are grouped in wider bins.
The final bin is defined even wider by construction; we tested using higher resolution bins for older populations and found that the machine could not accurately distinguish between different aged populations above $\sim 3 \; \mathrm{Gyr}$.

Before training any of our machine learning methods we first split the data in to training (72\%), validation (8\%) and test (20\%) sets.
We take care to perform any optimisation, be that normalisation of the features or hyperparameter optimisation, solely on the training (+ validation) data.

%% file: methods_ml.tex
\subsection{Machine Learning Methods}
\label{sec:ml_methods}
We implement two different learning algorithms: Extremely Randomised Trees (ERT) and Convolutional Neural Networks (CNN). Using two different methods provides an additional means of evaluating the performance through comparison.

\subsubsection{Convolutional Neural Networks}
\label{sec:cnn}

\begin{figure*}
	\includegraphics[width=\textwidth]{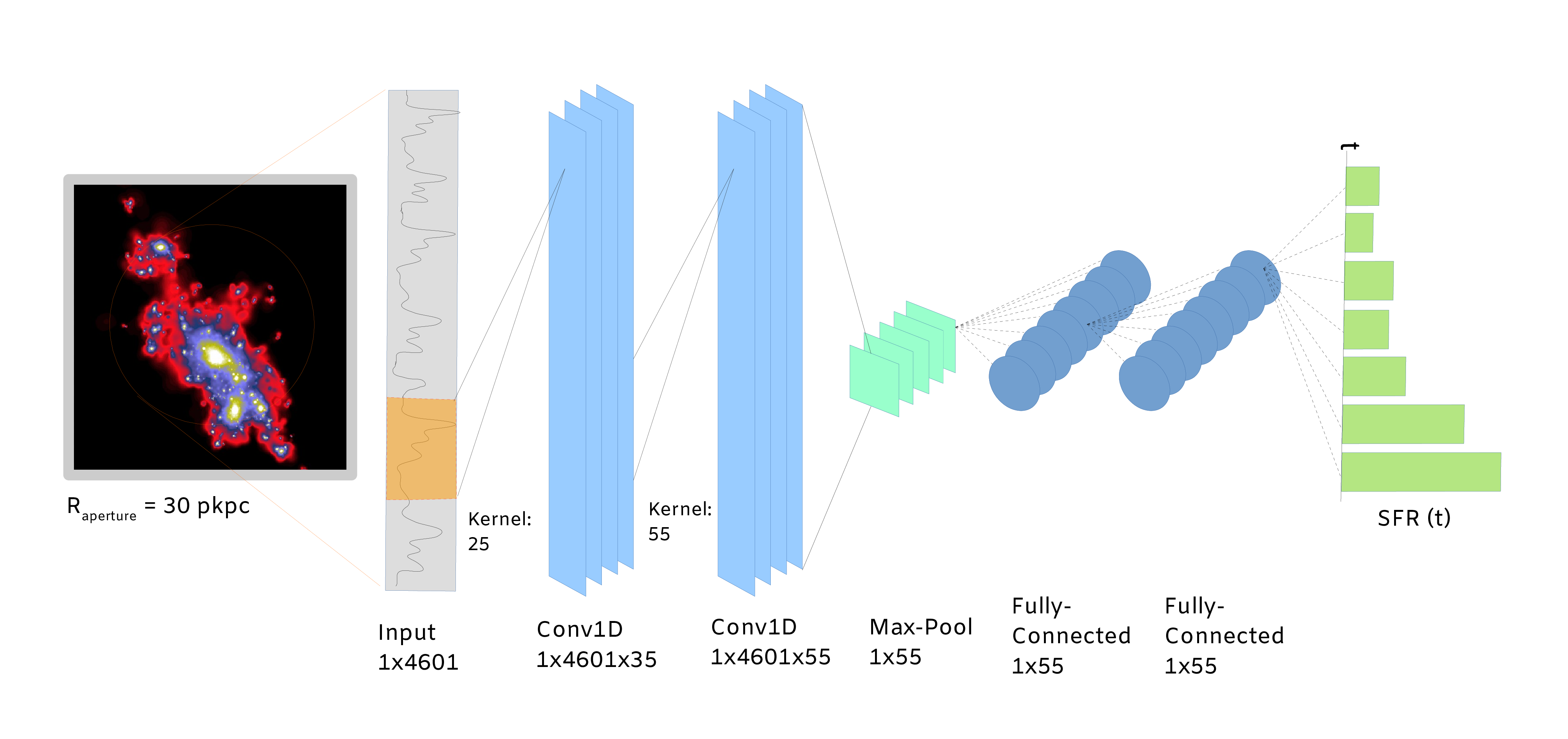}
    \caption{The CNN architecture, described in detail in \sec{cnn}.}
    \label{fig:cnn_architecture}
\end{figure*}

Convolutional Neural Networks (CNNs)\footnote{For further background see \cite{gu_recent_2018,kiranyaz_1d_2019,fan_selective_2019}} are growing in popularity in many areas of Astronomy, typically as a means of analysing 2D image data (\textit{e.g.} \citealt{tuccillo_deep_2017, petrillo_finding_2017}), and have been shown to perform remarkably well, with prediction accuracies in classification tasks approaching human level \citep{flamary_astronomical_2016, fabbro_application_2018}.

Our CNN architecture was inspired by the work of \cite{fabbro_application_2018}, who use the python version of Keras \citep{chollet2015keras} to apply the technique to 1D stellar spectra.
We make a number of modifications, as well as a systematic hyperparameter search given our training features.
The basic structure (shown in \fig{cnn_architecture}) uses two convolutional layers, the first applied directly to the one dimensional input spectral features, the latter applied to the outputs of the first layer.
The convolution operation essentially shares information between neighbouring pixels, allowing the network to identify spatial correlations in feature space, such as gradients and emission / absorption lines; tiered convolutional layers allow the model to learn higher order relationships.
The output of the second convolutional layer is then fed in to a max-pooling layer, which takes the maximum from each feature map generated from the convolutional layers, significantly reducing the dimensionality (from $1\times4601\times55$ to $1\times55$); this leads to faster training and reduced overfitting.
Finally, the output of the pooling layer is fed in to a traditional fully-connected neural network, where each neuron in a given layer is connected to every neuron in the subsequent layer.
We tested different configurations, from shallow and wide (few layers, many neurons in each layer) to deep and narrow (many layers, few neurons in each layer), and settled on the former.
The convolution and pooling layers together can be thought of as the \textit{feature extraction} part of the network, and the fully-connected layers perform regression on these features.

The network weights are initially set randomly, then updated through iterations of forward and back propagation utilising the Adam optimizer \citep{kingma_adam:_2014}.
We minimise Symmetric Mean Absolute Percentage Error (SMAPE; see \sec{loss_function}) as the target loss function.
The network is trained in epochs; during each epoch the training data are fed in batches (the batch size being a free parameter), and once all training galaxies have been used the trained model is evaluated on the validation set. This gives a validation score, that is used to decide when the training has converged, and to prevent overfitting.
During training we monitor the validation loss after each epoch and reduce the learning rate if it has plateaued, or stop training altogether if the improvement is below some threshold after a given number of epochs (early stopping), to prevent overfitting.

Optimising the network architecture is notoriously difficult due to the flexibility available in the network configuration.
However, once the general architecture has been decided, there are further optimisations that can be made to higher level hyperparameters that can lead to significant improvements.
We use \textsc{hyperas}\footnote{\href{https://github.com/maxpumperla/hyperas}{https://github.com/maxpumperla/hyperas}} to optimise a subset of these parameters: the number of filters and size of the kernel in each convolutional layer, and the number of neurons in the fully connected layers.
Hyperas utilises Tree-structured Parzen Estimators (TPE), which, after an initial random search, sequentially approximates the performance of hyperparameters based on previous measurements, building a likelihood based model \citep{bergstra2011algorithms}.

\subsubsection{Extremely Randomised Trees}
Ensemble decision tree algorithms aggregate the results of multiple trained decision trees in order to produce a single prediction, and can be applied to both classification or regression tasks. Since decision trees are computationally inexpensive to train, the training of ensembles does not lead to a significant performance penalty, and can be simply parallelised.
Extremely Randomised Trees \cite[ERT;][]{geurts_extremely_2006} is one such ensemble approach that has been successfully used in a wide range of Astronomy domains (\textit{e.g.} \citealt{kamdar_machine_2016,cohn_approximations_2018}). It is similar to the popular Random Forest (RF): during training of a RF, a subset of $K$ features is randomly chosen during each split, which reduces the correlation between trees where there are features with a strong correlation with the predictors.
ERT also perform this same feature space sampling, but add a further level of randomness by making non-deterministic split choices

We use the implementation of ERT provided in \textit{scikit-learn} \citep{scikit-learn}, with grid search cross validation to optimise the following hyperparameters: minimum samples in a split, minimum samples in a leaf, and maximum nodes in a leaf. This optimisation is done solely on the training set during each training procedure. For ERT, the full training set (training + validation) is used during training and optimisation.

\subsubsection{Loss Functions}
\label{sec:loss_function}
During model training and evaluation, the fit is assessed through a particular \textit{loss function}.
Typical loss functions include the mean absolute percentage error (MAPE) and the mean squared error (MSE), with the mean taken over all of the output predictors.
Both of these loss functions are inappropriate when applied to star formation histories sampled from a reasonably wide range of final stellar masses.
For example, the MSE leads to large penalties for histories with high SFH normalisation, whilst lower mass galaxies with a lower SFH normalisation are not penalised to the same degree despite similar percentage errors in their predictions.
On the other hand, it is not possible to calculate percentage errors for zero valued bins.

We would ideally like a loss function that acts as a percentage error, in order not to penalise high mass galaxies, but returns reasonable results for zero valued bins. We use a variation of Symmetric Mean Absolute Percentage Error (SMAPE),
$$\mathrm{SMAPE} = \left[ 2 \times \frac{\Sigma_{b} \, |\mathrm{Y}_{b}^{\mathrm{true}} \,-\, \mathrm{Y}_{b}^{\mathrm{pred}}|}{\Sigma_{b} \, (\mathrm{Y}_{b}^{\mathrm{true}} \,+\, \mathrm{Y}_{b}^{\mathrm{pred}})}\right] \times 100\%\,\,,$$
where $\mathrm{Y}_b$ is the star formation rate in bin $b$. The value of SMAPE is bounded between $0\% < \mathrm{SMAPE} < 200\%$, but acts as a true percentage error in the low error regime.
This point statistic can be used as both a reasonably unbiased loss function within the CNN, and as an evaluation of the fit.

%% file: methods_sims.tex
\subsection{Cosmological Simulations}
\label{sec:sims}

\begin{figure}
	\includegraphics[width=\columnwidth]{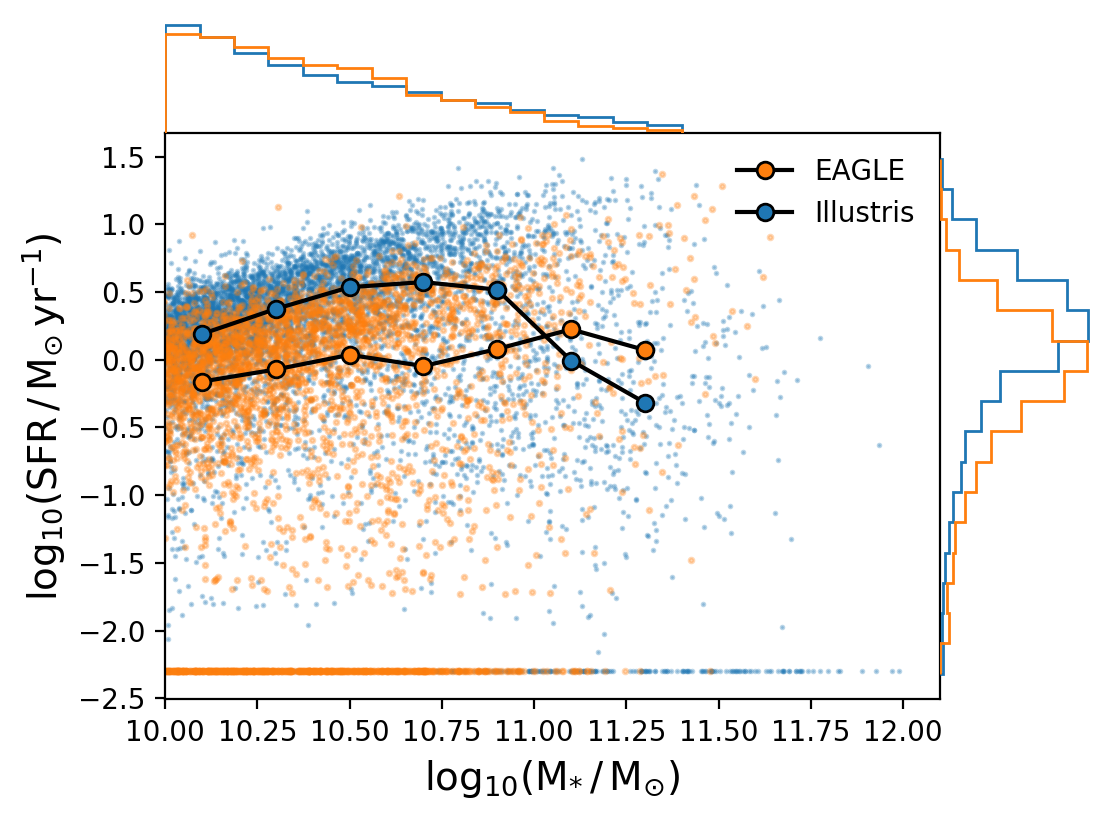}
    \caption{The $M_{*}$ - SFR relation, or star-forming sequence, at $z = 0.1$ for the selected Illustris and EAGLE galaxies. The scatter shows individual objects, and the median relation with $1\sigma$ spread is over-plotted. SFR is calculated using the integrated mass of stars formed in the last 100 Myr within a 30 pkpc aperture. Galaxies with zero recent SFR are plotted at $10^{-2.3} \; \mathrm{M_{\odot} \, yr^{-1}}$ for clarity. The histograms at the top and right of the plot show the normalised number counts as a function of stellar mass and SFR, respectively. EAGLE and Illustris show contrasting behaviour in this parameter plane.}
    \label{fig:mstar_sfr_sim_comparison}
\end{figure}

We use two cosmological hydrodynamic simulations, EAGLE \citep{schaye_eagle_2014, crain_eagle_2015} and Illustris\footnote{Galaxy and particle information for Illustris were obtained from the online API, \url{http://www.illustris-project.org/data/}} \citep{vogelsberger_introducing_2014, genel_introducing_2014}, which have both been run on large comoving volumes, tens of megaparsecs on a side, producing tens of thousands of galaxies at $z = 0$.
EAGLE\footnote{Galaxy and particle information for EAGLE were obtained from the public database, \url{http://icc.dur.ac.uk/Eagle/database.php} \protect\citep{mcalpine_eagle_2016,the_eagle_team_eagle_2017}} uses a modified version of the Smoothed Particle Hydrodynamics (SPH) code \textsc{GADGET 3} \mbox{\citep{springel_simulations_2005}}, whereas Illustris uses the moving-mesh code \textsc{AREPO} \citep{springel_e_2010}.
The typical gas element mass in each simulation is $\sim 10^{6} \, M_{\odot}$; below this mass scale physical processes cannot be modelled self consistently, so subgrid prescriptions are used to handle processes such as radiative cooling, star formation, stellar evolution, star formation feedback, black hole seeding, and AGN feedback.
Each hydrodynamic solver handles shocks and instabilities differently, but on the whole the choice of solver does not have a large effect on global galaxy properties; it is in the subgrid models that significant differences between the simulations are most apparent \citep{somerville_physical_2015}.

By using two different simulations we are able to evaluate how our algorithms generalize, by training them on a single simulation then testing its performance on another. We can then assess whether we are learning the \textit{intrinsic} relationship between galaxy SEDs and their SFHs, rather than learning about the relationship in a particular simulation.

Both EAGLE and Illustris have been shown to agree reasonably well with observed stellar mass and star formation rate distribution functions at low redshift, though there are still discrepancies both between the simulations and with the observations.
For example, EAGLE fits the low mass end of the Galaxy Stellar Mass Function (GSMF), but underestimates the normalisation at intermediate masses around the knee of the GSMF ($\sim 5 \times 10^{10} \, M_{\odot}$), whereas Illustris overestimates both the low mass and high mass number densities, but shows good agreement around the knee \citep{schaye_eagle_2014, genel_introducing_2014}.
Even greater discrepancies between the simulations can be seen in the distribution of specific Star Formation Rate ($\mathrm{sSFR = SFR} \,/\, M_{*}$) as a function of stellar mass, which in EAGLE shows a relatively flat relation up to $M_{*} \,/\, M_{\odot} \sim 10^{10}$, which then falls by $\sim 0.8$ dex; this agrees with the observations, but the normalisation is $\sim 0.3$ dex lower at all but the highest stellar masses \citep{schaye_eagle_2014}.
In contrast, Illustris remains flat out to $M_{*} \,/\, M_{\odot} \sim 10^{11}$ \citep{sparre_star_2015}; Illustris galaxies with Milky Way-like masses exhibit higher SFRs compared to EAGLE.

Such differences are to be expected due to the complexity of physical processes to be modelled at a large range of scales, and their resolution is a key goal of research in the field.
However, confusingly, the photometric colour distributions in both simulations have been shown to be in relatively good agreement with observations at low redshift over the same mass range \citep{trayford_colours_2015, vogelsberger_introducing_2014}.
This inconsistency, between the intrinsic physical properties and the predicted photometric distributions, is due to differences in the choice of SED modelling assumptions, particularly the magnitude of the dust correction.

Both simulations assume a Chabrier IMF, but adopt different cosmological parameters; Illustris assumes WMAP9 \citep{hinshaw_nine-year_2013}, EAGLE Planck13 \citep{planck_collaboration_planck_2014}, however these differences are expected to have negligible impact on the resulting galaxy distribution functions.

\subsubsection{Measurement Aperture}
\label{sec:sim_aperture}

A significant proportion of the stars in massive galaxies are located within an extended halo surrounding the central stellar concentration. These stars tend to be older, are often accreted from other systems through interactions, and therefore have a different SFH from those in the centre, which leads to spatial gradients in physical and observed stellar properties.
Both the integrated luminosity and the colour of a galaxy are therefore sensitive to the measurement aperture, and in order to facilitate comparison with observations similar apertures should be used when generating synthetic SEDs.
Unfortunately, this relies on the simulations having realistic spatially resolved star formation histories, which has not been extensively tested, and is also subject to resolution issues for small apertures.
We use a spherical 30 kpc aperture centred on the gravitational potential minimum, which has been shown to yield similar masses to a Petrosian aperture typically used in photometric observational studies \citep{schaye_eagle_2014}.
All quoted stellar properties ($M_{*}$, $\mathrm{SFR}$, SFH, etc.) are taken from the star particles within this aperture, and synthetic spectra are generated using only these star particles (see \sec{spectra}); this must be taken in to account when comparing to observational studies (see \sec{obs_methods}).

\subsubsection{Galaxy Selection}
\label{sec:sim_selection}

We select all galaxies from each simulation at $z = 0.1$ with $\mathrm{M_{*} \,/\, M_{\odot}} > 10^{10}$, which gives 3687 and 6473 galaxies for EAGLE and Illustris, respectively.
The large offset is an unfortunate result of the difference in the GSMF normalisation between the simulations at the high mass end.
\fig{mstar_sfr_sim_comparison} shows the distribution of our selections on the $\mathrm{M_{*} \,-\, SFR}$ plane.
The normalised histogram at the top shows the distribution of stellar masses; the steepness of the GSMF in both simulations means that their are many more low mass galaxies than high.
Since these low mass galaxies dominate our training sample, we expect to see a degree of overfitting to such galaxies with respect to their less numerous high mass counterparts. We explore this in more detail in \sec{results}.

Illustris shows a steeper star-forming sequence relation than EAGLE and a higher normalisation between $10 < \mathrm{log_{10}(M_{*} \,/\, M_{\odot})} < 11$, but above this Illustris galaxies have lower SFRs.
Such significant differences in training and test data present a unique challenge for machine learning methods, where the accuracy on unseen data is usually poor, and as such represents a robust test of our method.

%% file: methods_spectra.tex
\subsection{Synthetic Spectra}
\label{sec:spectra}

The composite spectrum of a galaxy in each simulation is dependent upon the physical properties and spatial distribution of the stars, gas and black holes.
We ignore the AGN contribution, which we do not expect to have a great effect on the optical emission.

The pipeline for generating spectra detailed in this section is contained within the \textsc{spectacle} module, available at \href{https://github.com/christopherlovell/spectacle}{https://github.com/christopherlovell/spectacle}.

\subsubsection{Intrinsic Spectra}

\begin{figure}
	\includegraphics[width=\columnwidth]{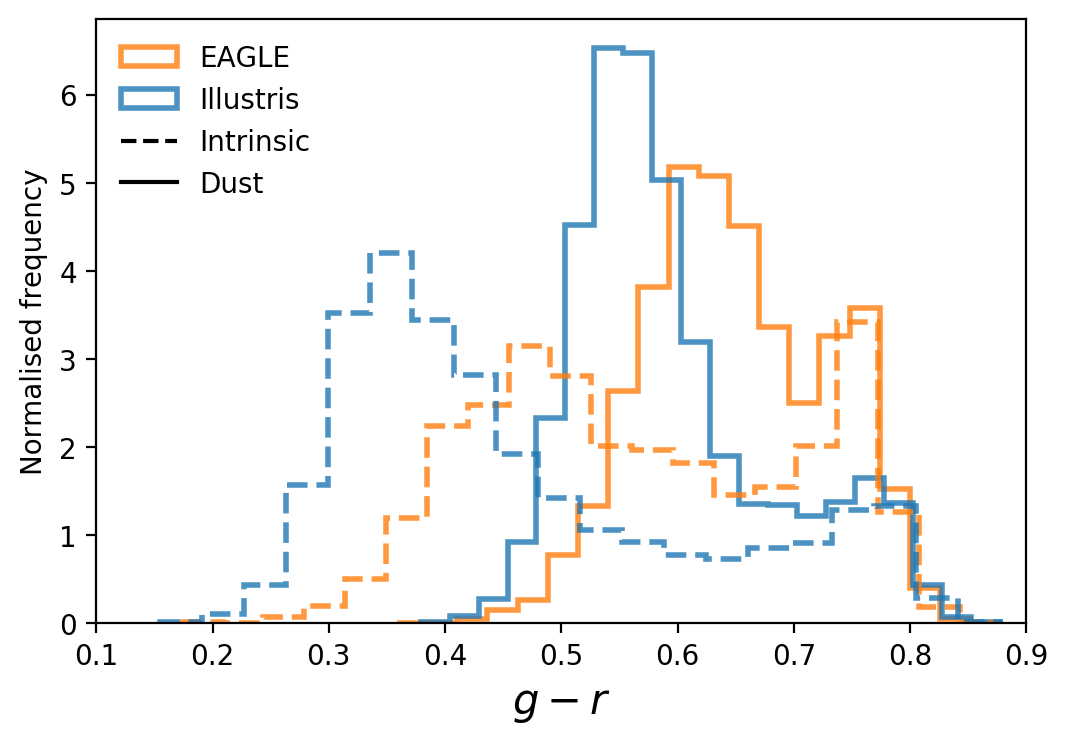}
    \caption{$g-r$ colour distribution for the EAGLE and Illustris simulation selections. Dashed lines show the intrinsic distributions (including the nebular contribution); solid lines show the dust-attenuated distributions. The dust model leads to a significant reddening of the blue population in both simulations.}
    \label{fig:g_r_distribution}
\end{figure}

\begin{figure}
	\includegraphics[width=\columnwidth]{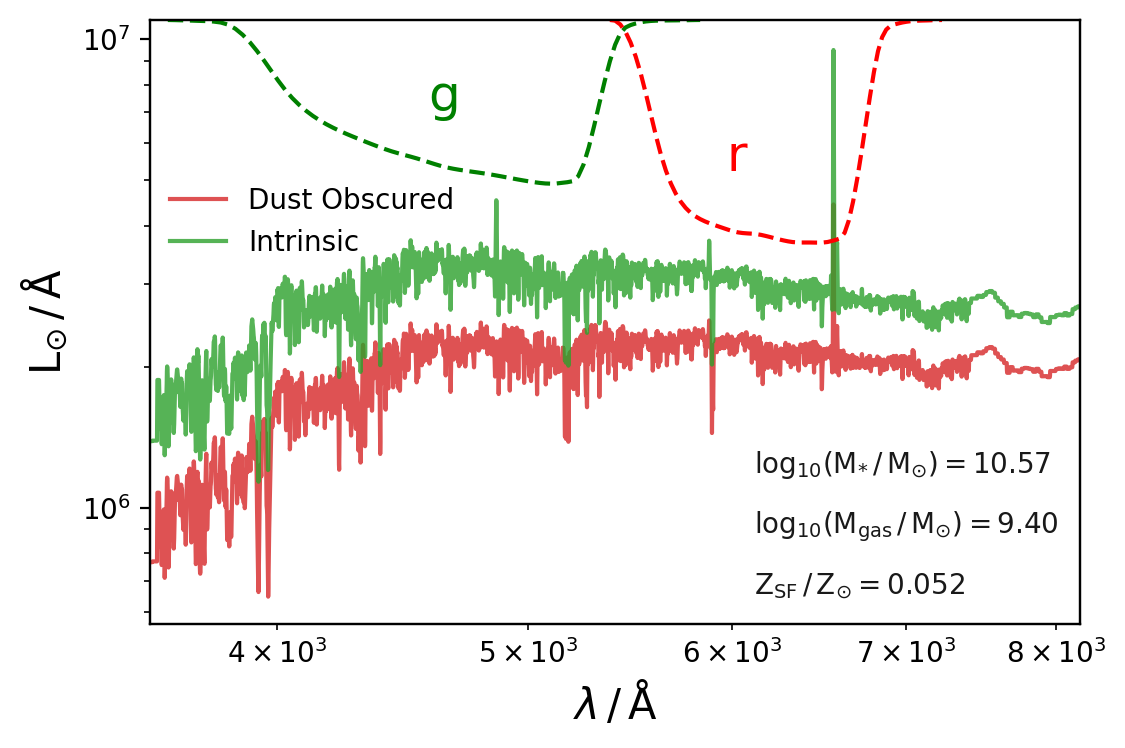}
    \caption{Intrinsic (green) and dust-obscured (red) spectrum for an example galaxy from the Illustris simulation. The $g$ and $r$ filter curve responses are shown at the top of the plot.}
    \label{fig:example_spectra}
\end{figure}

We generate intrinsic spectra by treating each star particle as a simple stellar population (SSP).
We then generate an SED for each SSP using the Python implementation of the Flexible Stellar Population Synthesis (FSPS) code \citep{conroy_propagation_2009, conroy_propagation_2010, dan_foreman_mackey_2014_12157}.
The SED of each SSP is dependent on its age and metallicity, normalised by its initial stellar mass.
Each stellar particle in the simulations is approximately two orders of magnitude more massive than typical star forming regions; a single young star particle can therefore significantly affect the predicted colours of a galaxy. In order to mitigate this artificial Poisson scatter we resample the recent star formation using a similar technique to that used in \cite{trayford_colours_2015}.
We take each star particle younger than 100 Myr and split it into ten thousand new particles with ages sampled uniformly within this interval, and the mass of the original particle equally distributed between the resampled particles.

Young stellar populations ionise their surrounding gas, leading to nebular line and continuum emission. This emission can dominate photometric fluxes, as well as being responsible for the majority of optical emission lines \citep{anders_spectral_2003, reines_importance_2010,wilkins_theoretical_2013}.
\cite{byler_nebular_2017} use the photoionization code \textsc{Cloudy} to model the expected nebular emission from young FSPS SSPs self-consistently; these templates are provided in python-FSPS.
They assume a covering fraction of unity for stellar populations with age $t < t_{\mathrm{esc}}$, where $t_{\mathrm{esc}} = 10^{7}$ years.

We define the `intrinsic' emission as including the nebular contribution. \fig{g_r_distribution} shows the intrinsic $g-r$ colour distribution for EAGLE and Illustris.
\fig{example_spectra} shows the example intrinsic emission for an Illustris galaxy; strong nebular line emission and absorption are clearly visible.

\subsubsection{Dust Attenuated Spectra}
\label{sec:method_dust}

\begin{figure}
	\includegraphics[width=\columnwidth]{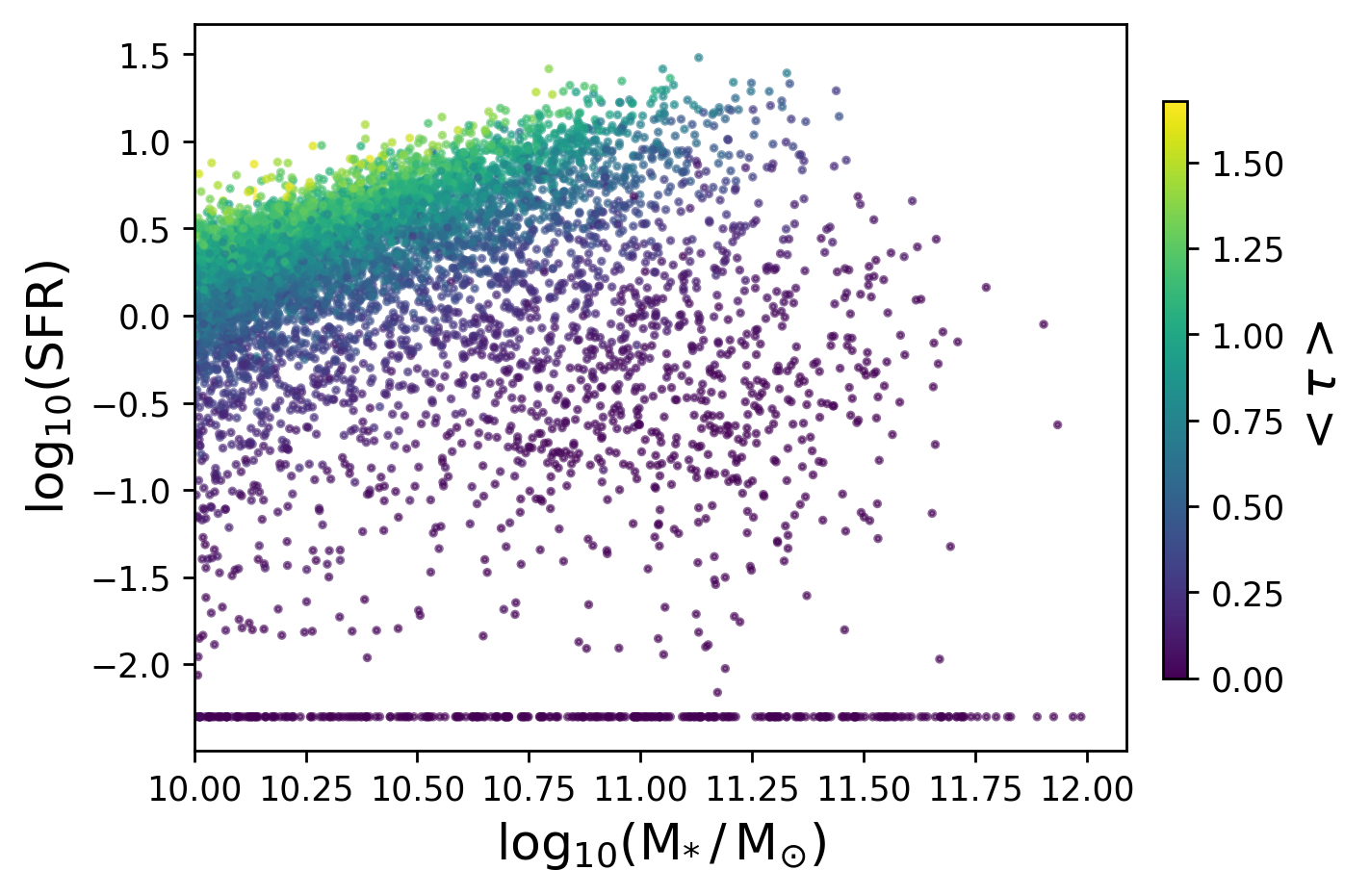}
    \caption{The star-forming sequence for the Illustris sample, coloured by the average attenuation over the whole galaxy (${<\tau> \,=\, - \mathrm{log}(F^{\mathrm{dust}}_{\lambda} \,/\, F^{\mathrm{int}}_{\lambda})[\lambda=5500 \; \mbox{\AA}]}$). Gas-rich, star-forming galaxies experience greater attenuation than gas-poor galaxies at the same stellar mass.}
    \label{fig:attenuation_illustris}
\end{figure}

We use information from the models on the mass and metallicity of star forming gas to provide a self-consistent, physically motivated prescription for the dust attenuation.
Our model assumes a simple screen, ignoring the dust distribution geometry.
The transmission $T$ at a wavelength $\lambda$ for a particle of age $t$ is given by
\begin{align}
T(\lambda, t)  = \mathrm{exp} \left[ - \tau(t) \left( \frac{\lambda}{\lambda_{\nu}} \right) ^{-1} \, \right] \;\;,
\end{align}
where $\tau$ is the optical depth at wavelength $\lambda_{\nu}$.
The optical depth is dependent on the age of the stellar particle; all particles are subject to a constant screen due to dust in the ISM, but young particles, which still reside within their birth clouds, are subject to a further transient attenuation component,
\begin{align*}
t \leq t_{\mathrm{disp}} \,&:\; \tau = \gamma \, \tau_{\,\mathrm{cloud}} + \gamma \, \tau_{\,\mathrm{ISM}}\\
t \geqslant t_{\mathrm{disp}} \,&:\; \tau = \gamma \, \tau_{\,\mathrm{ISM}} \;\;.
\end{align*}
Both $\tau_{\,\mathrm{ISM}}$ and $\tau_{\,\mathrm{cloud}}$ can be fixed constants ($\gamma = 1$), or linked to other properties of the galaxy. We link the optical depth to the metallicity and mass of cold, star forming gas:
\begin{align}
\gamma = \frac{Z_{\,\mathrm{SF}}}{Z_{\,\mathrm{Z14}}} \left( \frac{M_{\,\mathrm{SF}}}{M_{*}} \frac{1}{\beta} \right) \;\;,
\end{align}
where $Z_{\,\mathrm{SF}}$ is the mass-weighted star forming gas phase metallicity, and the mass dependence is encapsulated in the ratio of $M_{\,\mathrm{SF}}$, the total mass of star forming gas, to $M_{*}$, the stellar mass.
These are both normalised to the respective Milky Way values: $Z_{\,\mathrm{Z14}} = 0.035$
\footnote{This is taken from the $M_{*} - Z$ relation expression in \cite{zahid_universal_2014} evaluated at the Milky Way stellar mass, and converted to relative solar metallicities assuming $12 + \mathrm{log_{10}(O/H)_{\odot}} = 8.69$ \citep{allende_prieto_forbidden_2001}.}, and $\beta = 0.1$.
We use $\tau_{\, \mathrm{cloud}} = 0.67$, $\tau_{\, \mathrm{ISM}} = 0.33$, $t_{\, \mathrm{disp}} = 10 \, \mathrm{Myr}$ and $\lambda_{\nu} = 5500 \,\mbox{\text\AA}$, as used in both EAGLE and Illustris studies \citep{trayford_colours_2015,genel_introducing_2014}.
This approach produces a physically motivated attenuation, where gas rich spirals are subject to higher attenuation than gas poor ellipticals with identical stellar mass.
This can be seen in \fig{attenuation_illustris}, which shows the star-forming sequence for the Illustris selection, coloured by the mean attenuation.

\fig{example_spectra} shows the dust-obscured spectrum for an example Illustris galaxy. The high relative gas mass and star-forming gas phase metallicity leads to significant attenuation.
\fig{g_r_distribution} shows the distribution of $g-r$ colour for the dust attenuated spectra. Dust leads to a reddening of the blue population, shifting the peak by $\Delta (g-r) \sim +0.2$ in both simulations, but the location and normalisation of the red population in both cases is generally unaffected; this is expected since these intrinsically red systems are generally gas poor, and experience lower attenuation.

\subsubsection{Artificial Noise}
\label{sec:method_noise}
In order to further increase the realism of our synthetic spectra we add artificial noise at a given signal to noise (SN) level. We use a fiducial value of SN = 50, and test the effect of increased SN on our predictions in \sec{model_training}.

For each spectrum we can take multiple realisations of the noise. This can be useful in two ways: it can increase our training set, and it can prevent the model from overfitting to a single noisy realisation by providing multiple noise-added spectra for a given SFH.
We explore the effect of using multiple noisy realisations on our model training in \sec{model_training}.

\subsubsection{Wavelength Grid}
\label{sec:wavelength_grid}

We restrict the wavelength coverage to that approximately covered by the SDSS DR7 release (see \sec{obs_methods}), and resample \citep[flux preserving;][]{carnall_spectres:_2017} on a fixed logarithmically-sampled wavelength grid.
This gives a final fixed input wavelength grid, $3572 \leqslant \lambda \,/\, \mbox{\text\AA} \leqslant 8173$, with resolution $\lambda \,/\, \Delta \lambda = 5570 \;\; (\lambda = 4500 \, \mbox{\text\AA})$.

%% file: results.tex
\section{Results}
\label{sec:results}

We first train both Extremely Randomised Trees (ERT) and Convolutional Neural Network (CNN) models on our \textsc{EAGLE} and Illustris training samples (80\% of the data). All plots in this section show predictions when applied to galaxies in the respective test sets (20\% of the data).

\subsection{Training \& Testing}
\label{sec:model_training}

\subsubsection{Learning Curves}

\begin{figure}
	\includegraphics[width=\columnwidth]{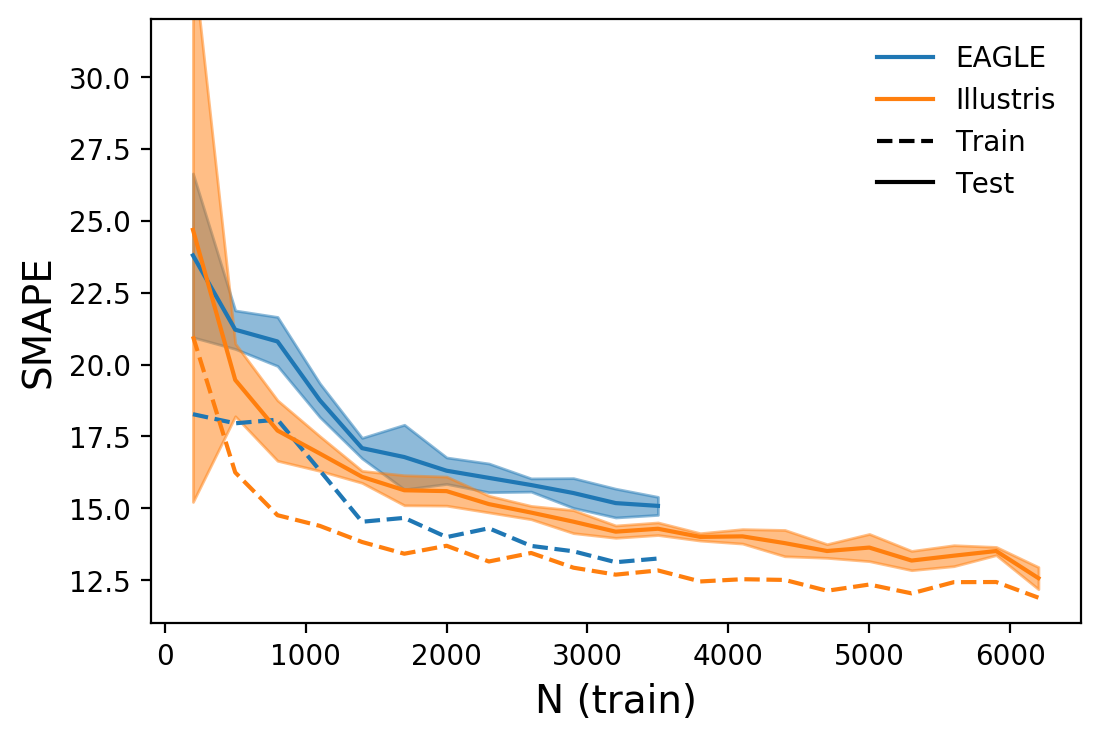}
    \caption{Learning curves, showing the SMAPE as a function of input training data size, from CNN trained on dust attenuated spectra from both Illustris and EAGLE. Multiple samples without replacement are drawn from the full training set, and the median SMAPE on the training and test sets are shown as the dashed and solid lines, respectively. The shaded region showing the 1$\sigma$ spread in the test SMAPE.
		}
    \label{fig:learning_curves}
\end{figure}

Learning curves show the improvement in test score as a function of training set size, which provides information on the convergence of the model.
Decreasing scores suggest that a larger training set would lead to a better fit, whereas a plateau suggests that the training has converged and no further improvement can be obtained from additional training data.
A large gap between the training error and the test error would indicate overfitting, or poor generalisation properties. \fig{learning_curves} shows learning curves for dust attenuated spectra from Illustris and EAGLE.
We perform 6-fold cross validation to estimate the scores and present their median.
The EAGLE learning curve is still falling at 3500 samples, which suggests that the model is yet to converge. The Illustris learning curve, in comparison, appears to have plateaued at $\sim \, 5500$ samples, though a larger training set is needed to confirm this.
As a result, we concentrate on the converged Illustris model for the time being (we will return to the EAGLE training set later, both in conjunction with the Illustris training data, and as an independent test set for the Illustris trained model).
The gap between the training and test errors in both EAGLE and Illustris is small, which suggest negligible overfitting.
The EAGLE model has slightly higher SMAPE at fixed N than Illustris, but it is unclear what specific differences in the simulation modelling lead to this; a possible explanation is the higher gas-phase metallicity in EAGLE at fixed stellar mass compared to Illustris, which will contribute to greater dust attenuation, obscuring the underlying relationship between the SFH and the spectra more in EAGLE than Illustris.

\subsubsection{Method comparison}

\begin{figure}
	\includegraphics[width=\columnwidth]{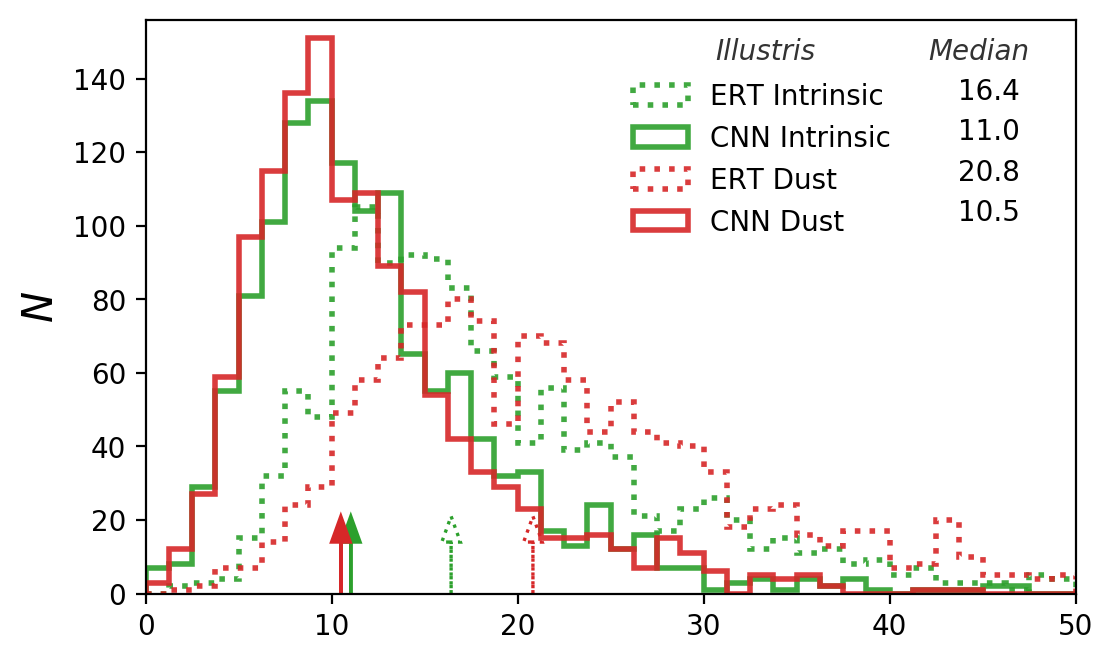}
	\includegraphics[width=\columnwidth]{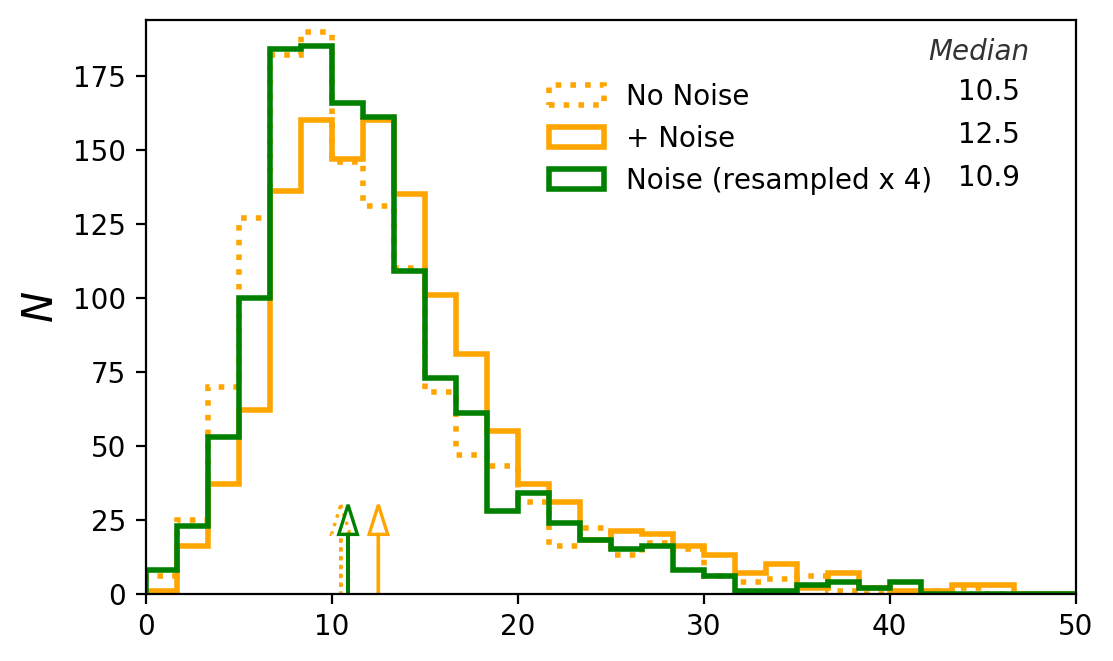}
	\includegraphics[width=\columnwidth]{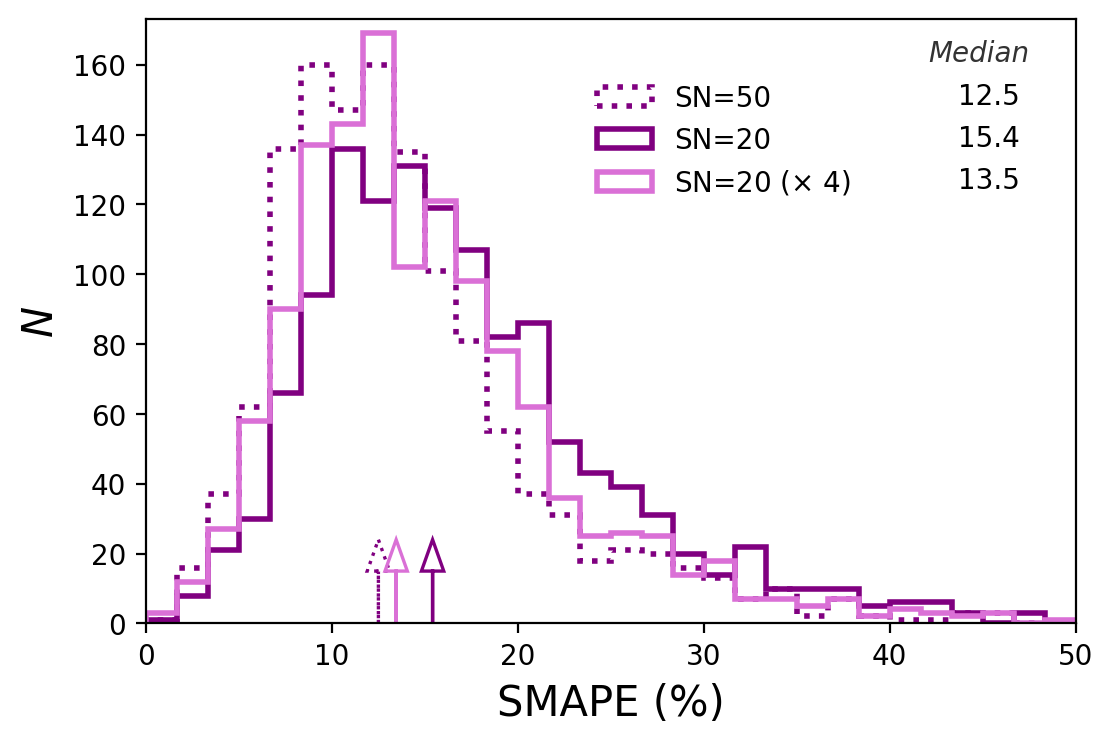}
    \caption{SMAPE distributions for the Illustris simulation, with different learning algorithms and spectral modelling. The median of each distribution is shown by the arrows, and quoted in the legend. \textit{Top:} ERT (dashed) and CNN (solid) models trained on intrinsic (green) and dust-obscured (red) spectra.
		\textit{Middle:} CNN model trained on dust-obscured spectra (dashed), with added noise (solid, yellow), and with noise resampled $\times$ 4 (solid, green).
		\textit{Bottom:} CNN model trained on dust-obscured spectra with added noise at SN=50 (dashed, purple), SN=20 (solid, purple), and with noise resampled $\times$ 4 at SN=20 (solid, pink).}
    \label{fig:smape}
\end{figure}

\begin{figure*}
	\includegraphics[width=\textwidth]{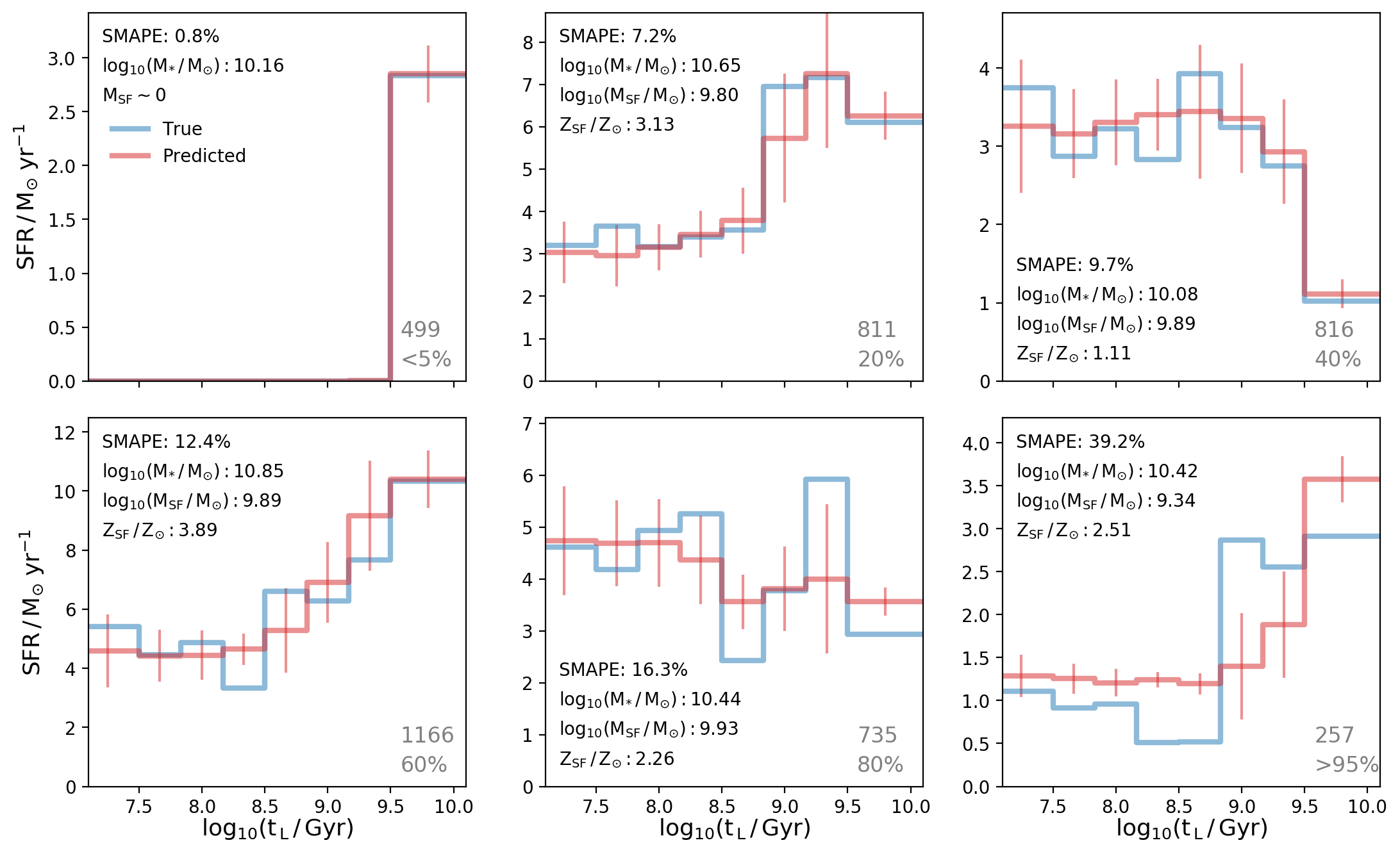}
    \caption{Six example SFHs from the Illustris test set (blue), alongside fits to the dust-obscured spectra (red). The examples are selected with a range of SMAPE scores, 0.8-40.0\%, from top left to bottom right. Errors are a combination of observational and modelling errors (see \sec{errors}). Each panel shows the galaxy index and the approximate SMAPE score percentile in the bottom right, as well as the $z=0.1$ stellar mass, star-forming gas mass and star-forming gas metallicity.}
    \label{fig:panel_plot}
\end{figure*}

Returning to the Illustris data in isolation, the top panel of \fig{smape} shows the distribution of SMAPE scores, for both ERT and CNN and for dust obscured and intrinsic spectra, evaluated over the entire Illustris test set.
The median SMAPE for the CNN is significantly lower than for ERT for both intrinsic and dust obscured features.
This is due to the CNN's ability to share local information between neighbouring pixels, whereas ERT treats each pixel as an isolated feature.
We also find that the median SMAPE for dust obscured spectra with ERT is significantly higher than that for intrinsic spectra, however for the CNN this difference is negligible.
Dust introduces additional degeneracies between the spectral features and the underlying SFH, so it interesting that the CNN is capable of overcoming these.

We choose to focus on the CNN performance in the rest of the paper.

\subsubsection{Model Results with Noise}

As mentioned in \sec{method_noise} we add noise to our simulated spectra with a fiducial value of SN=50.
The middle panel of \fig{smape} shows the SMAPE distribution for a model trained with this added noise, and as expected the noise leads to an increase in the median SMAPE of 2\%.
However, we can re-sample the noise for each synthetic spectrum multiple times.
Using a multi-resampled training set leads to a reduction in the SMAPE; we tested different numbers of resamples, and found that the improvement in SMAPE plateaus at 4.
The SMAPE distribution using this 4 times resampled feature set is shown in the middle panel of \fig{smape}; the median SMAPE is much lower than for the single noise-realisation feature set (10.9\%).
This suggests that the negative effect of the noise, that obscures the relationship between the spectra and the SFH, is overcome by the positive impact of the larger, more generalisable training set.

We expect the prediction accuracy to decrease as the noise level is increased.
To test this, we used a SN = 20, shown in the bottom panel of \fig{smape}. This leads to an increase in the SMAPE of 2.9\% compared to the fiducial SN = 50.
However, as in the lower noise case, resampling the noise $4 \; \times$ leads to an improvement of 1.9\% in the median SMAPE over the single-realisation model.
We quote results using the SN=50, 4 $\times$ resampled spectra in the rest of this section, unless otherwise noted.

\subsubsection{Example Fits}

In order to illustrate the SFH fits we show six examples from the Illustris test set in \fig{panel_plot}.
We show predictions for a range of SMAPE scores as evaluated on the dust attenuated SEDs.
The top left panel shows one of the best fits, the next four panels show fits around the 20th, 40th, 60th and 80th percentiles of the SMAPE distribution, and finally the bottom right panel shows one of the worst fits.
The errors on the fit in each bin are taken from the observational and model errors combined in quadrature (see \sec{errors}).

\subsubsection{Parameter Correlations}
\label{sec:param_correlations}

\begin{figure}
	\includegraphics[width=\columnwidth]{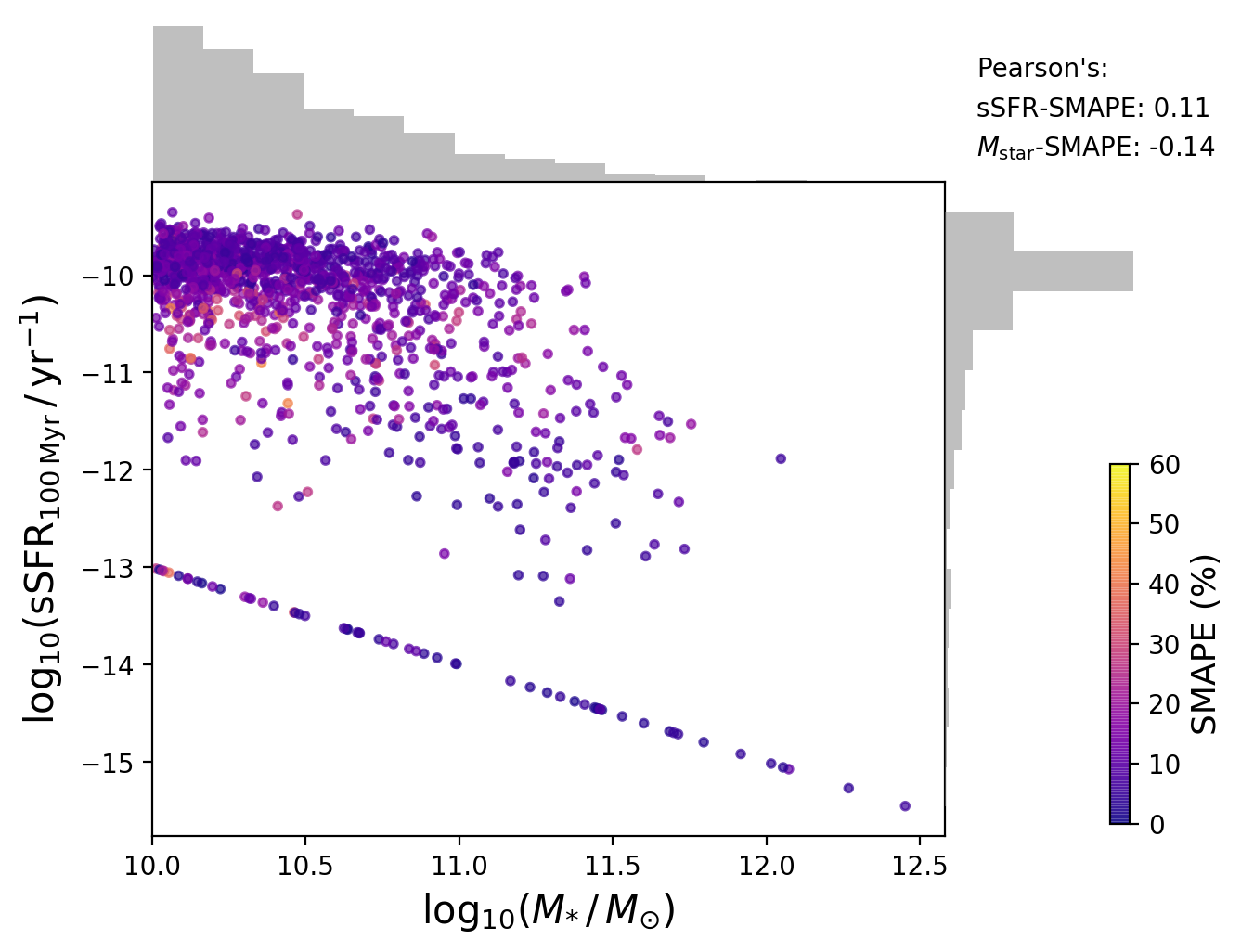}
  \includegraphics[width=\columnwidth]{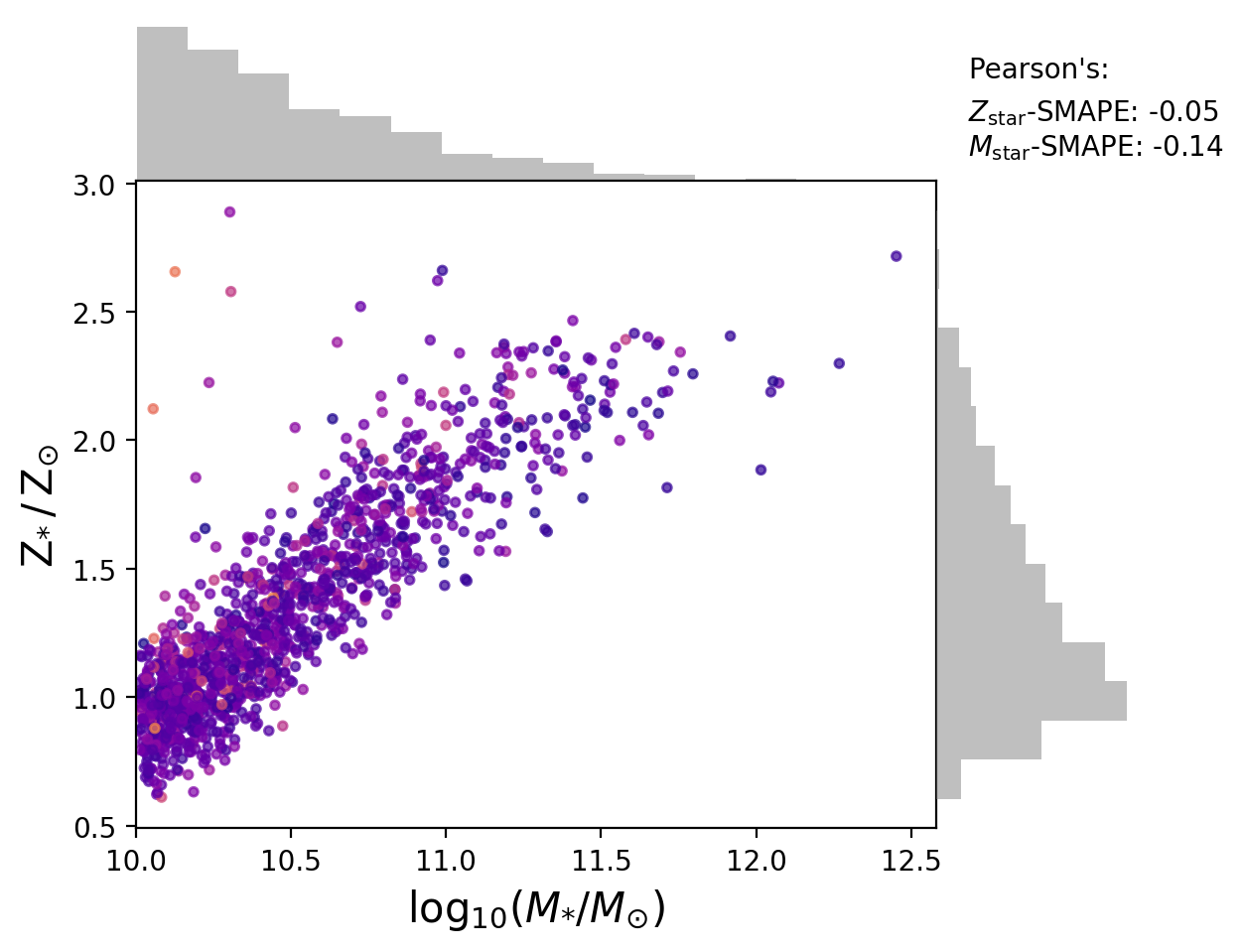}
    \caption{Parameter correlations with SMAPE for the predictions on the Illustris test set, using the intrinsic spectra. The pearson's correlation coefficient between each parameter and SMAPE is shown in the top right. The grey histograms above and to the right of each axis show the distribution of the given parameter. \textit{Top:} stellar mass - SFR relation. SFR is calculated as the integrated mass in stars formed in the last 100 Myr. \textit{Bottom:} stellar mass - stellar metallicity relation.}
    \label{fig:smape_mstar}
\end{figure}

As mentioned in \sec{sim_selection}, we preferentially select low mass galaxies due to the steepness of the GSMF.
It is therefore important to investigate any correlation of the quality of fit with stellar mass, to evaluate any overfitting to low mass galaxies.
The top panel of \fig{smape_mstar} shows the distribution of Illustris test galaxies on the stellar mass - SFR plane, coloured by SMAPE on the predicted histories from the dust-attenuated model.
In order to quantify any trend of SMAPE with our galaxy parameters we calculate the Pearson's correlation coefficient,
$$\rho = \frac{\mathrm{cov}(P,SMAPE)}{\sigma_{P}\sigma_{SMAPE}} \;\;,$$
where $P$ is the given parameter, $\mathrm{cov}$ is the covariance between the parameter and SMAPE, and $\sigma$ is the standard deviation of the respective quantity.
There is no significant correlation between stellar mass and SMAPE ($\rho = -0.14$), nor between specific-SFR and SMAPE ($\rho = 0.11$).

The well known age-metallicity degeneracy in the optical can also obscure the underlying SFH \citep{worthey_comprehensive_1994}.
The bottom panel of \fig{smape_mstar} shows the stellar mass - metallicity distribution, for the Illustris test galaxies, coloured by SMAPE on the intrinsic model.
There is no significant correlation between stellar metallicity and SMAPE ($\rho = -0.05$).
This may be due to the relatively low resolution of the SFHs, reducing the confusion between bins.

%% file: testing.tex
\subsection{Testing Across Simulations}

\begin{figure}
	\includegraphics[width=\columnwidth]{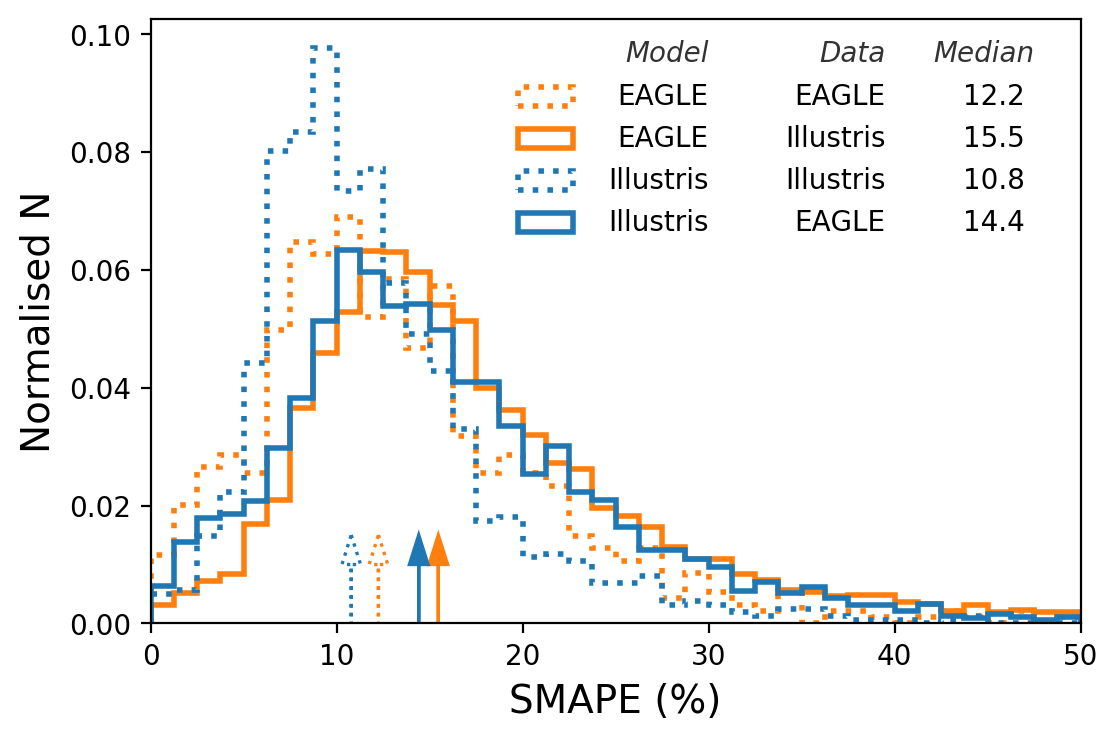}
    \caption{The normalised SMAPE distribution for the inter-sim (solid) and within-sim (dashed) test sets, for dust-attenuated spectra. The median of the distribution is shown by the arrow on the x-axis, and quoted in the legend. Despite being trained on very different data, the SMAPE is low in both inter-sim cases.}
    \label{fig:smape_intersim}
\end{figure}

\begin{figure}
	\includegraphics[width=\columnwidth]{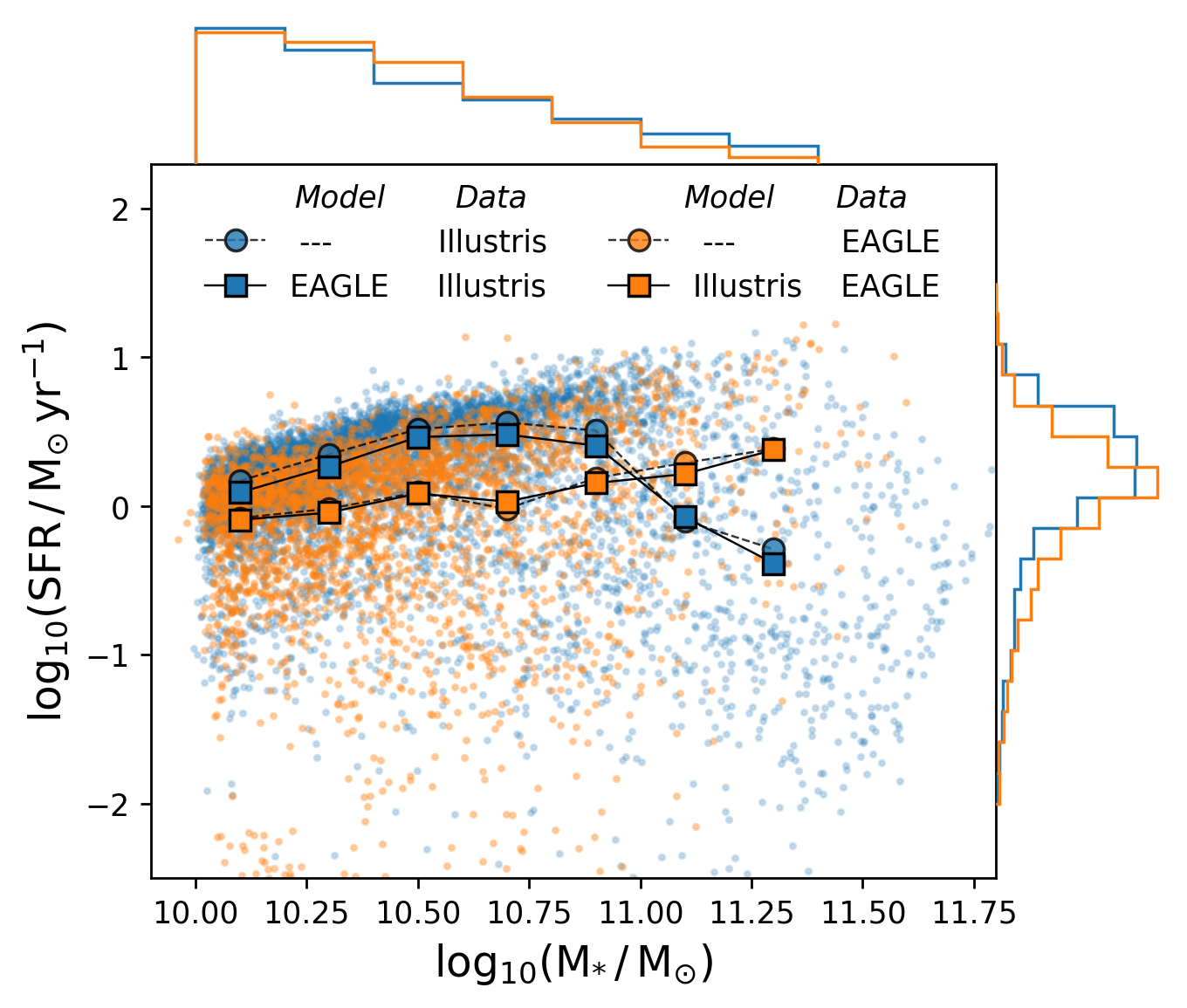}
    \caption{The predicted star-forming sequence for the intersim results. We estimate the present day SFR from the normalisation in the latest SFH bin, corresponding to a timescale of approximately 30 Myr, and the total mass from the SFH combined with an age-dependent recycling fraction. Each model prediction, shown with the square points and solid lines, recovers the original star-forming sequence, shown by the circular points and dashed lines, despite being trained on SFHs corresponding to a different SFR-$M_{*}$ relationship.}
    \label{fig:intersim_sfs_dust}
\end{figure}

Further uncertainty is introduced by our choice of modelling assumptions, such as the training simulation, SPS model, intrinsic SED pipeline and dust model. Of these we expect the choice of training simulation to lead to the greatest bias.
To estimate the uncertainty introduced we test a model trained assuming some simulation training data on another model trained assuming different simulation training data.
This procedure demonstrates how well each model generalises.

\fig{smape_intersim} shows the SMAPE error when our CNN is trained and tested on different simulations, using dust-obscured spectra with 4 $\times$ resampled noise.
Since the latter testing simulation is not included in any of the training, the full galaxy sample can be used for testing; we plot the normalised distributions to aid comparison.
For models trained on both EAGLE and Illustris the median SMAPE for the intra-sim results is higher than within-sim.
The errors are still reasonably good in all intra-sim cases, despite the significant differences in the simulations used for the training and testing data.

Another way of testing whether the model is overfitting is to plot the predicted distribution of galaxies on the stellar mass-star formation rate plane.
We have already seen in \fig{mstar_sfr_sim_comparison} that both simulations exhibit very different behaviour in this space, and might expect a model that has overfit to a particular simulation to recover the distribution from its training data.
\fig{intersim_sfs_dust} shows that this is not the case: each model recovers the star-forming sequence of the new input data.

\begin{figure}
	\includegraphics[width=\columnwidth]{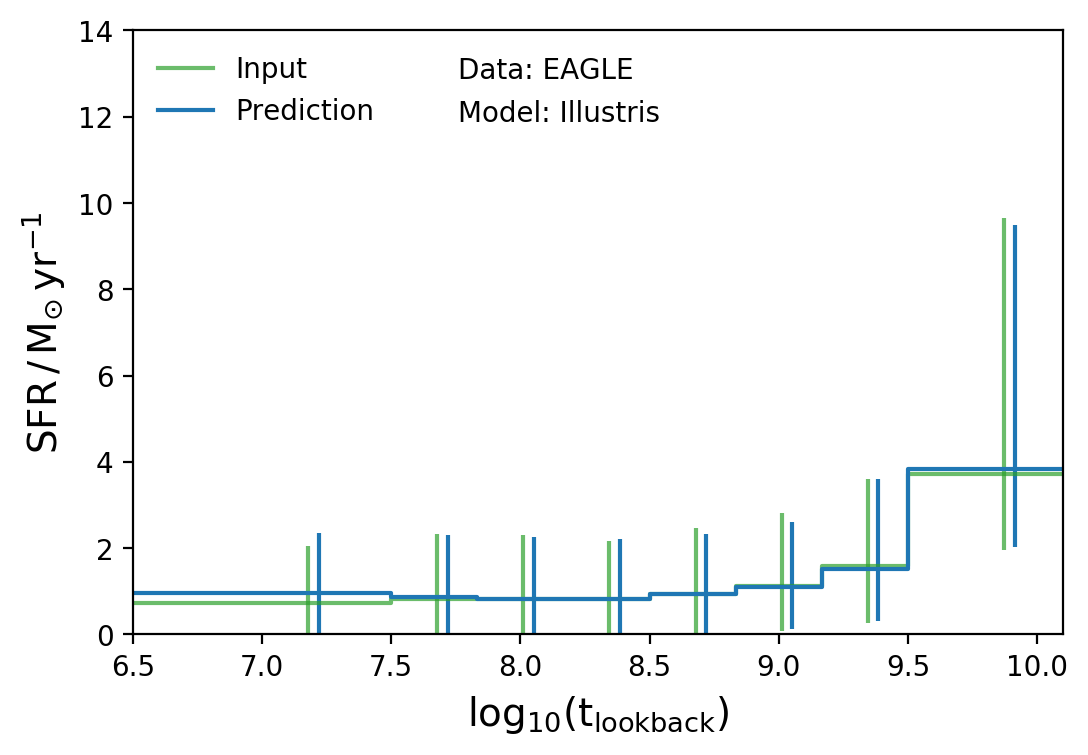}
	\includegraphics[width=\columnwidth]{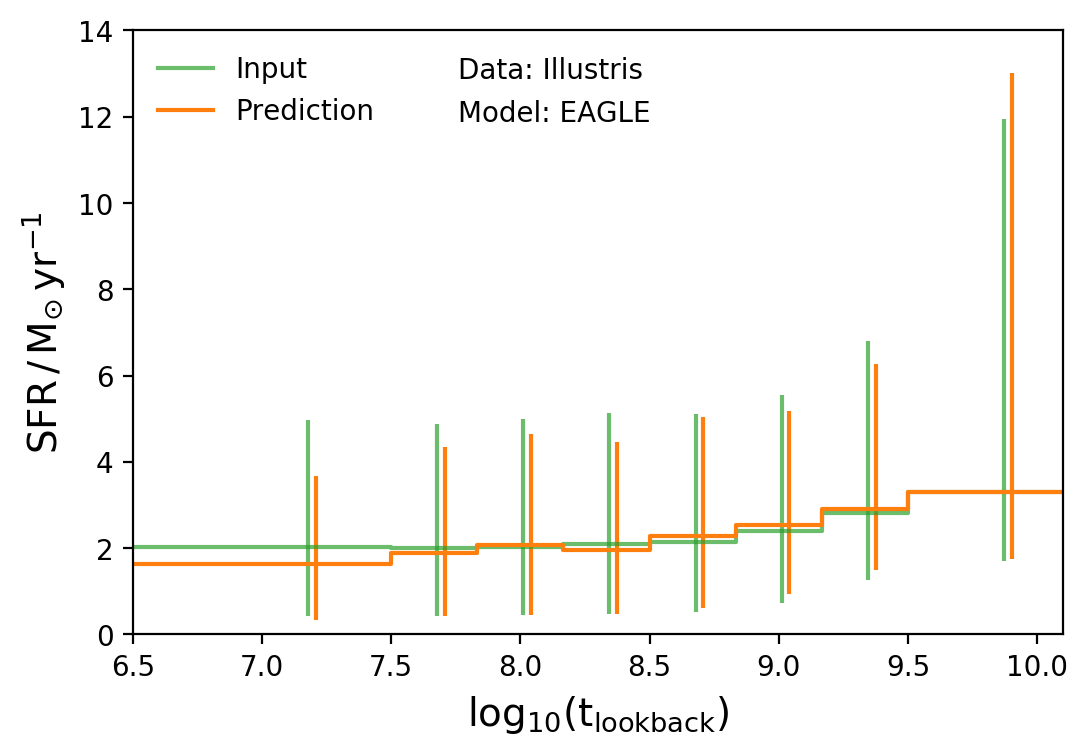}
    \caption{The median SFH and $16^{\mathrm{th}}-84^{\mathrm{th}}$ percentile spread in each bin for the input data (green) and the intersim prediction (orange for the EAGLE mode, blue for the Illustris model). The distribution of predicted SFHs is recovered well in both cases.
		}
    \label{fig:ensemble_average_intersim}
\end{figure}

Whilst these integrated and point-in-time properties are recovered accurately, the shape of the SFH, and the distribution of SFHs, may still be incorrectly predicted.
To test this we show in \fig{ensemble_average_intersim} the median and $16^{\mathrm{th}}-84^{\mathrm{th}}$ percentile spread in each bin for the input data and the predictions.
The distribution of predicted SFHs is remarkably similar for both simulations throughout cosmic time.

%% file: errors.tex
\section{Error Estimates}
\label{sec:errors}

Our SFH predictions are subject to two main sources of uncertainty: those from errors in the spectra, which we refer to as \textit{observational} errors, and those from errors in the CNN fit, which we refer to as \textit{modelling} errors.
In this section we make estimates for the impact of these two sources of error, and combine them to give a total estimated error in each bin.

\subsection{Observational Errors}
\label{sec:spec_errors}

Errors in the observed SED will lead to uncertainty in the predicted histories.
The propagated error can be estimated in two ways: sample a number of noisy SEDs, predict the SFHs for each noise-added spectrum, and calculate the covariance matrix of the output, as in \cite{tojeiro_public_2009}, or treat the model as a vector valued function and evaluate the dot product of the Jacobian and the error spectrum, as demonstrated in \cite{fabbro_application_2018}.
Errors calculated with both procedures should give similar results since they are essentially evaluating the same input dependence; the former does this through Monte Carlo sampling, whereas the latter explicitly calculates the gradient of the predictors with respect to the features.

We implement the former approach, using the noise model described in \sec{method_noise}.
For each spectrum we add $N$ random realisations of each error spectrum to the input spectrum, and propagate each noise-added spectrum through our model to obtain a distribution of predicted histories.
From these the covariance matrix can be calculated,
$$ C_{ij} = \left\langle (x_{i} - \hat{x}_{i})(x_{j} - \hat{x}_{j}) \right\rangle \;\;,$$
where $x_i$ is the SFR in bin $i$ for a given realisation, and $\hat{x}_{i}$ is the mean SFR in that bin for all realisations.
The uncertainty in each bin is then $\sigma_{i} = \sqrt{C_{ii}}$.
We can also use $C$ to find the correlation matrix; we describe this in more detail, alongside examples, in \app{correlation_matrices}.

\begin{figure}
	\includegraphics[width=\columnwidth]{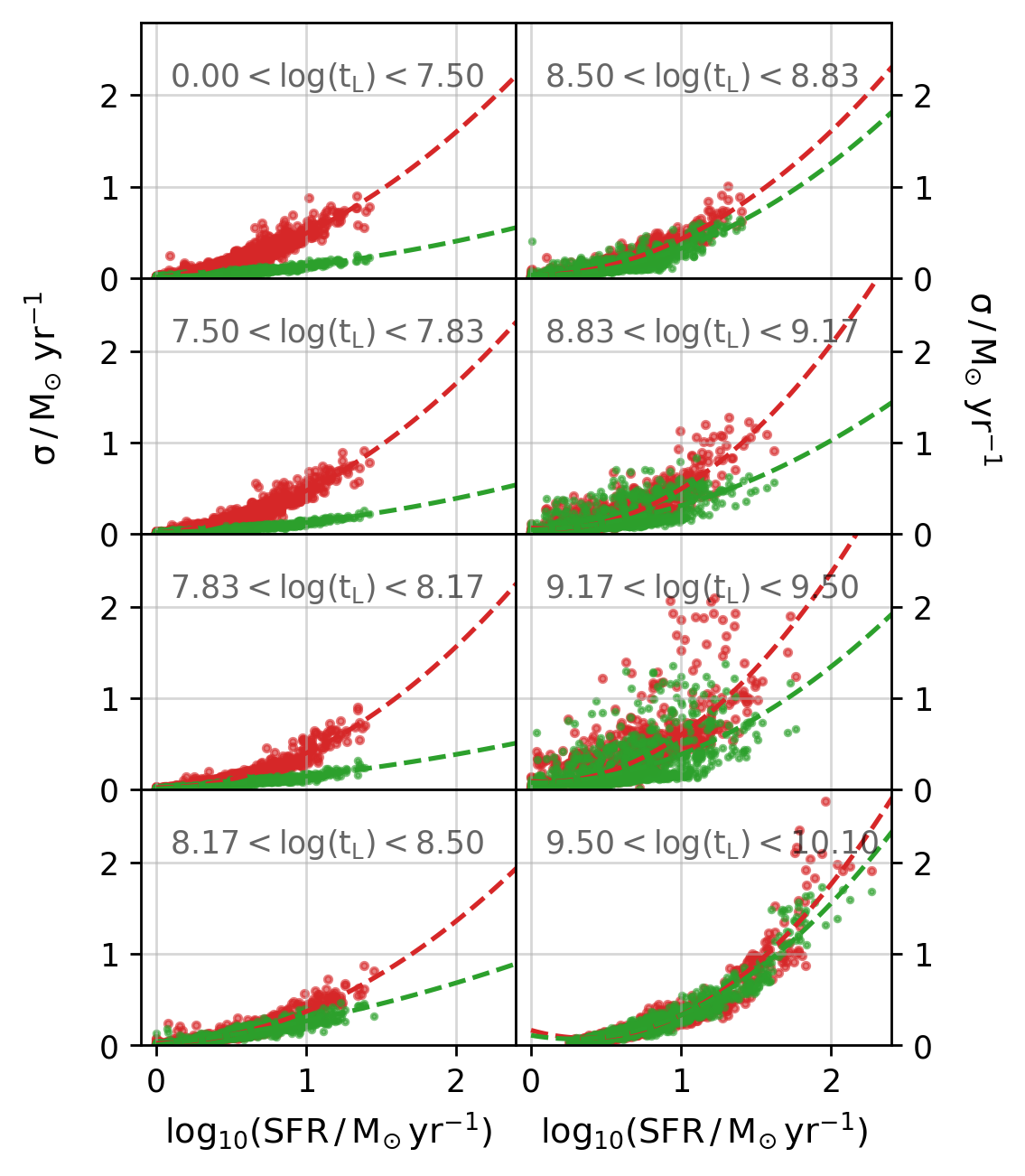}
    \caption{Observational errors (1$\sigma$) as a function of SFR in each bin, for intrinsic (green) and dust-obscured (red) spectra. Second order polynomial fits are shown as dashed lines. Observational errors are strongly dependent on the quantitative SFR, and are larger for dust-obscured spectra in recent bins.}
    \label{fig:spectra_errors_illustris}
\end{figure}

\fig{spectra_errors_illustris} shows the observational error in each bin as a function of SFR, for intrinsic and dust-obscured spectra.
The error is positively correlated with the quantitative value of the SFR.
In all but the oldest bin, the errors on dust attenuated spectra are larger than in the intrinsic case.
We fit second order polynomials to the $\sigma\;-\;\mathrm{log_{10}(SFR)}$ relation for each bin, which allows us to predict the observational error for arbitrary histories (fit parameters are quoted in \app{total_errors}).

\subsection{Modelling Uncertainties}
\label{sec:model_uncertainty}

\begin{figure*}
	\includegraphics[width=\textwidth]{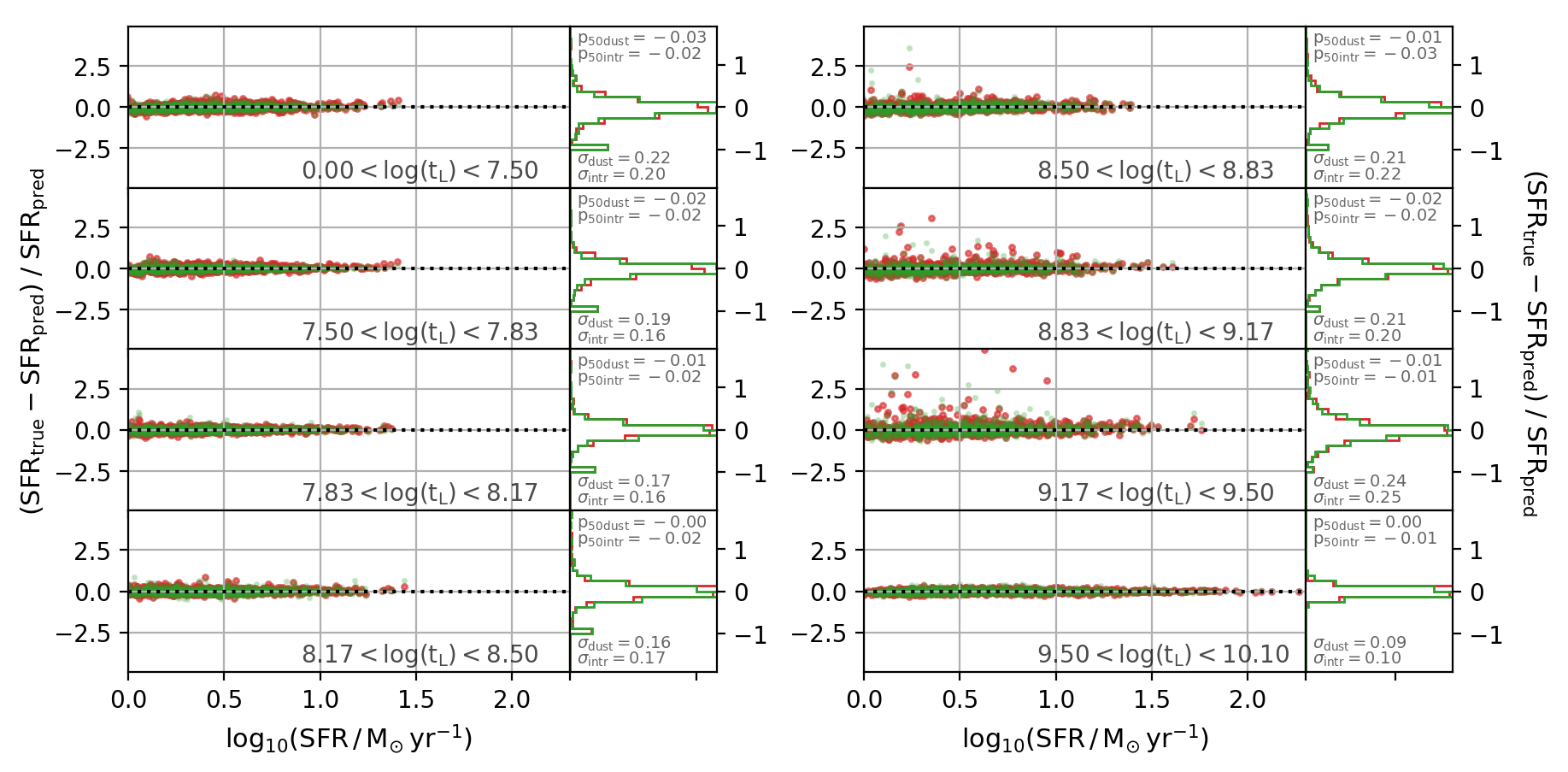}
    \caption{Fractional residuals between the true SFH and the predicted SFH for intrinsic (green) and dust attenuated (red) spectra from Illustris. The residuals are plotted as a function of the logarithm of the absolute star formation. The right panels show a one dimensional histogram of the distribuion of fractional residuals, with mean and 1$\sigma$ spread from a normal fit quoted in each panel.}
    \label{fig:fractional_residuals_by_mass_illustris}
\end{figure*}

There are a number of free parameters in our model pipeline, from the synthetic SED generation to the parametrisation of the dust model, to the free parameters of the CNN.
It is impractical to estimate the uncertainty on each parameter, however we can obtain an estimate of the propagated model uncertainty directly from the scatter of the residuals in predicted SFH.
The magnitude of the residual is SFR dependent in all bins; we account for this by dividing by the absolute predicted SFR in the bin to give the \textit{fractional residual}.
This single statistic can be used to estimate the model error for each galaxy, bin-by-bin, by multiplying by the predicted SFR.

\fig{fractional_residuals_by_mass_illustris} shows the fractional residuals between the predicted and the true SFR in each lookback-age bin as a function of the true SFR within that bin, along with normal fits to the marginalised distributions.

\subsection{Total Error}
We combine the observational and modelling errors to obtain the \textit{total} error by adding them in quadrature.
Since the error is dependent on the quantitative SFH in each bin we do not quote it, but provide fits to the observational error and fractional residual distributions in \app{total_errors}.
The modelling errors dominate the error budget for all bins, for an observational error SN = 50; we have tested up to SN = 20, and this remains the case.
\fig{panel_plot} shows the total uncertainties calculated using this method, for each example.

%% file: methods_obs.tex
\section{Observational Predictions}
\label{sec:obs_summary}

We apply the model to the SDSS DR7 Main Galaxy Sample (MGS)\footnote{obtained from the Data Archive Server, das.sdss.org} \citep{strauss_spectroscopic_2002, abazajian_seventh_2009}, which allows us to compare with \textsc{Vespa} \citep{tojeiro_recovering_2007, tojeiro_public_2009}, an SED fitting code for predicting SFHs that has been applied to this catalogue.
\textsc{Vespa} uses similar binned star formation histories to our method, allowing a like-for-like comparison between the two methods.
The level of agreement in predicted SFHs, or lack thereof, does not imply that either technique is more robust, but simply allows us to highlight the differences between our approach and an SED fitting approach.

\subsection{SDSS Selection}
\label{sec:obs_methods}

\begin{figure*}
	\includegraphics[width=\textwidth]{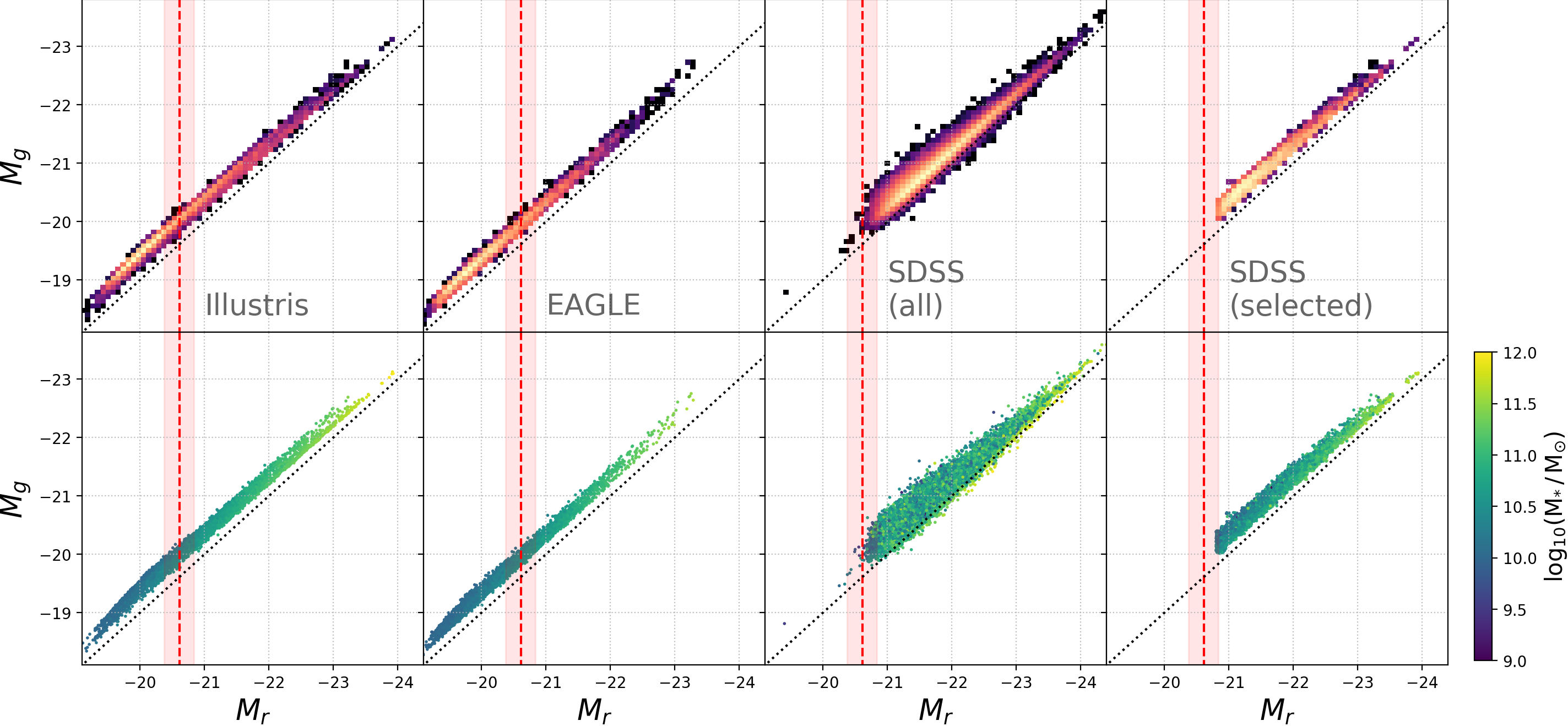}
    \caption{$g'$ and $r'$ magnitude distributions in EAGLE, Illustris, all SDSS galaxies, and our final magnitude- and mass-limited selection (left to right). The red dashed line in all panels shows the SDSS DR7 target magnitude limit $r'_{\mathrm{lim}}$ at $z = 0.1$. The red shaded region shows the extent of $r'_{\mathrm{lim}}$ for $0.09 \leqslant z \leqslant 0.11$. \textit{Top panels:} the number density. Scale is not consistent between panels. \textit{Bottom panels:} the stellar mass distribution. For the simulations this is the intrinsic stellar mass within the aperture. For SDSS this is the VESPA stellar mass estimates.}
    \label{fig:sdss_g_r_selection_mass}
\end{figure*}

We first selected all MGS galaxies in the redshift range $0.09 < z < 0.11$ where the redshift confidence was higher than 95\%, which gave 76812 objects. We then removed those galaxies whose rest-frame wavelength coverage, with bad pixels removed, did not cover our fixed wavelength grid (see \sec{wavelength_grid}), yielding 66245 galaxies.
Given our fixed wavelength grid we interpolated each spectrum \citep[flux preserving;][]{carnall_spectres:_2017}, de-redshifted and corrected for galactic extinction \citep{barbary_kyle_2016_804967} using the \cite{schlegel_maps_1998} galactic dust maps for each SDSS plate combined with the \cite{odonnell_rnu-dependent_1994} extinction curves ($R_{V} = 3.2$, where $R_{V} = A_{V} / E(B-V)$).

\subsubsection{Aperture Correction}
SDSS spectra are taken through a 3 arcsecond diameter fibre, which corresponds to $6 \; \mathrm{pkpc}$ at $z = 0.1$. In order to apply our model, trained on galaxy spectra generated using a 30 pkpc aperture intended to mimic a petrosian aperture, we chose to scale up the observed fluxes by the mean of the difference between the fiber and petrosian magnitudes in the \textit{observer} frame $g$ and $r$ bands (henceforth $g'$ and $r'$),
$$S = 10^{0.2 \,\times\, ([M^{\mathrm{fiber}}_{g'} - M^{\mathrm{petro}}_{g'}] + [M^{\mathrm{fiber}}_{r'} - M^{\mathrm{petro}}_{r'}])} \;\;,$$
where $S$ is the flux scaling factor.
After these corrections, the magnitude distribution on the $g - r$ plane of the selection at this stage can be seen in the top panel, third from left, of \fig{sdss_g_r_selection_mass}.
An alternative to scaling up the observational fluxes would have been to generate spectra from the simulations using a mock fibre aperture.
Unfortunately, as discussed in \sec{sim_aperture}, on these small scales numerical resolution effects become important.

\subsubsection{Colour Selection}
We then used rest frame $g$ and $r$ magnitudes to perform a 2D selection on $g$ and $r$ band magnitude simultaneously (without replacement), in order to match the same 2D distribution from the combined Illustris and EAGLE samples (see the first two panels of \fig{sdss_g_r_selection_mass}).
SDSS spectra have a target apparent magnitude limit\footnote{https://classic.sdss.org/dr7/} of $r' < 17.77$, which corresponds to an absolute magnitude of -20.61 at $z = 0.1$; a large proportion of our simulated galaxies lie below this threshold, so we are limited to matching the distribution above this constraint (as shown by the red dotted line in all panels of \fig{sdss_g_r_selection_mass}).
Selecting galaxies above this threshold with matched magnitudes gives us a sample of $10\,000$ galaxies.
It is clear from \fig{sdss_g_r_selection_mass} that the $g$ against $r$ distribution for SDSS galaxies deviates from 1:1 more so than the simulations, motivating the 2D selection.
The selection based on the simulation broadband magnitudes is to ensure that, when used as features for the model, the spectra remain `in-bounds' to some extent, i.e. are not outside the range of input training data.
It is true that we \textit{a priori} select observed galaxies with good spectral agreement with our simulations in a broad-band sense, however the details of the higher resolution spectra can still differ substantially.
We have tested that our models do not fail dramatically on out-of-bounds SDSS data, however a more thorough test with simulated out-of-bounds performance is left for future work.

In \app{t_sne} we show how t-SNE can be used to evaluate the synthetic gap between the synthetic and observed spectra.

%% file: obs_comparison.tex
\subsection{VESPA Star Formation Histories}
\label{sec:obs_vespa}

The \textsc{Vespa} SFH catalogue predicts star formation histories with varying resolution depending on the quality and completeness of the input data, with a maximum resolution of 16 bins, though a resampled SFH at this higher resolution is also provided. We use this resampled SFH throughout the comparison, though caution that this does not necessarily represent the best fitting history.
\textsc{Vespa} also provides predictions using the SPS models of both \cite{bruzual_stellar_2003} and \cite{maraston_evolutionary_2005}. The choice of model leads to significant differences in the predicted SFH, which highlights the effect of modelling choices on the inferred SFH.
We use the \textsc{Vespa} results that use BC03 models assuming a Chabrier IMF, whilst noting that these will not necessarily lead to consistent predictions compared to the more recent FSPS models used in our model training, and do not include nebular emission.
Using the more recent FSPS model is justified since the improved spectral modelling will lead to galaxies with more comparable intrinsic properties, such as the SFH, particularly since our selection is magnitude-matched to the SDSS sample.
We leave a comparison of the effect of SPS model choice to future work.

SDSS DR7 spectra are measured within fiber apertures of 3'' diameter.
\textsc{Vespa} SFHs are corrected for this by scaling the entire normalisation (\textit{i.e.} the mass in each bin) by the offset between the fiber and petrosian $z$-band magnitudes \citep{tojeiro_public_2009},
$$M_{*,fiber} = \frac{M_{*,total}}{10^{0.4(z_{f} - z_{p})}} \;\;.$$
where $z_{f}$ and $z_{p}$ are the fibre and petrosian $z$-band magnitudes, respectively\footnote{In \cite{tojeiro_public_2009} the equation for the stellar mass correction contains an error; it is reproduced here correctly}.

\subsection{SDSS Predictions}
\label{sec:obs_pred}

\begin{figure}
	\includegraphics[width=\columnwidth]{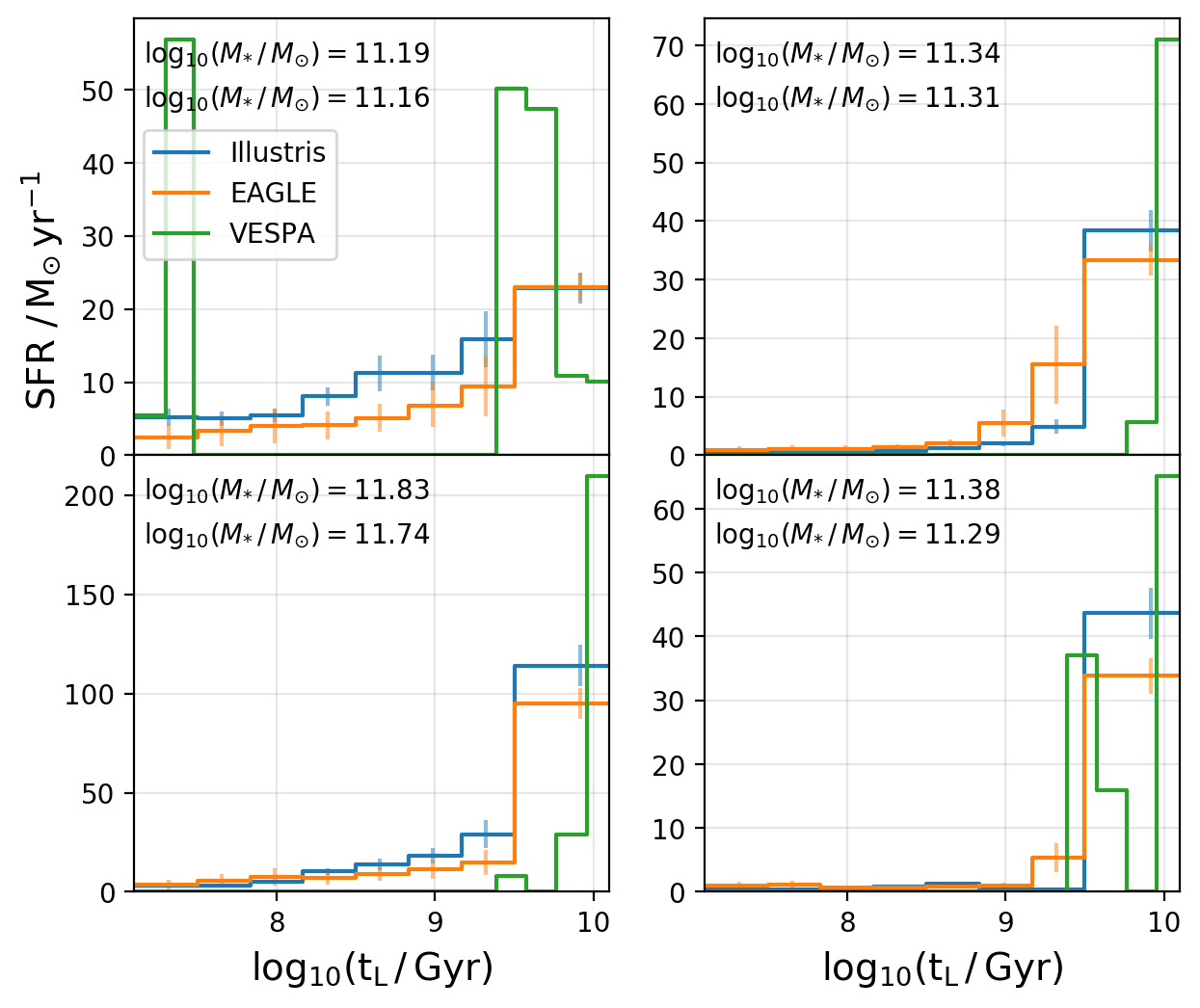}
    \caption{Four example SFHs from \textsc{Vespa}, alongside predictions for the same SDSS galaxies from the EAGLE and Illustris models (trained on dust-obscured spectra with noise, resampled $\times \;3$). We show histories with total predicted masses from the Illustris model closest to the estimated \textsc{Vespa} total masses. Uncertainties are estimated from the observational and modelling errors, described in \sec{errors}. Our models trained with EAGLE and Illustris predict similar shaped histories, with smoother evolution than \textsc{Vespa}.}
    \label{fig:SDSS_SFH_comparison}
\end{figure}

\begin{figure}
	\includegraphics[width=\columnwidth]{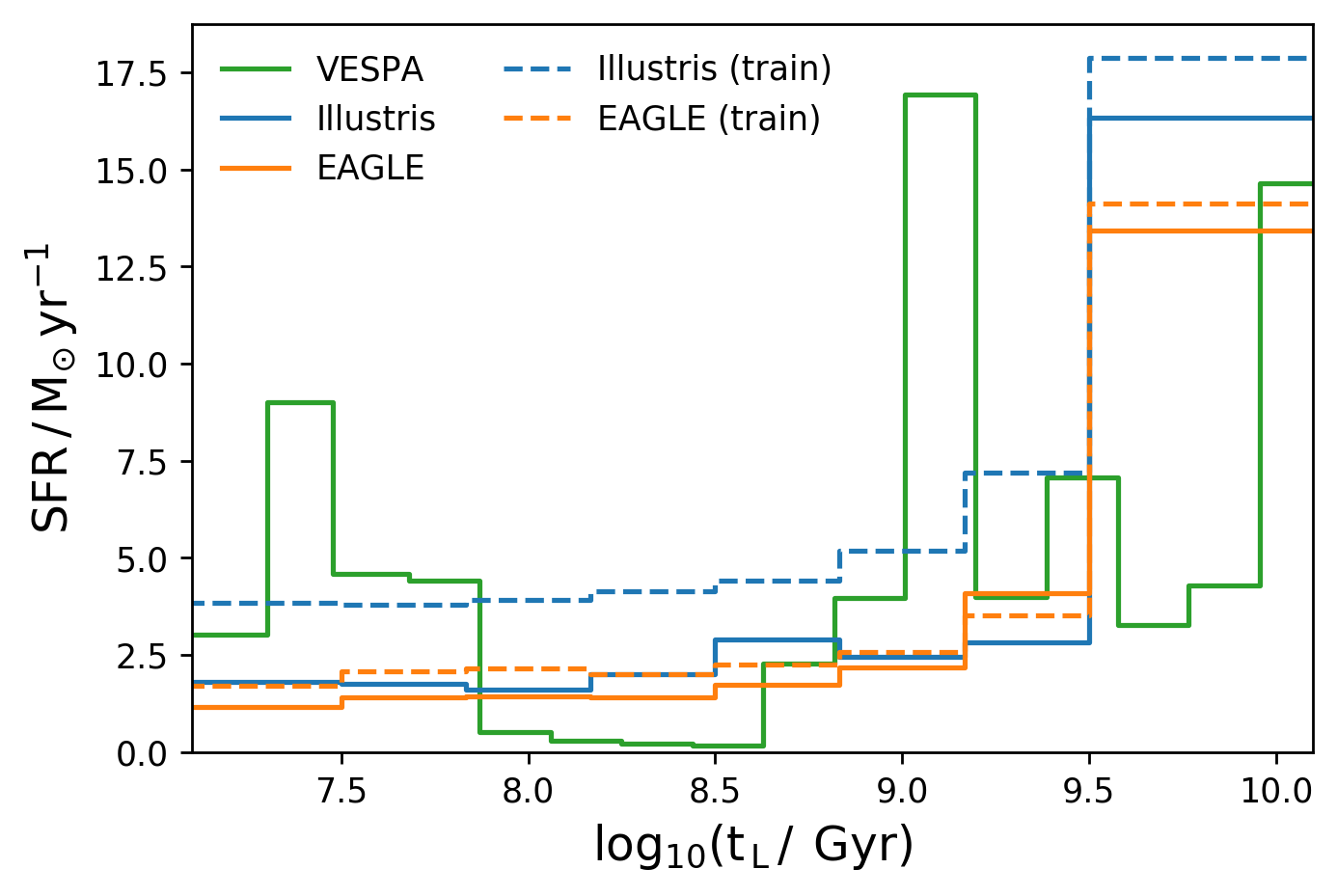}
    \caption{Mean predicted SFH for the SDSS selection from \textsc{Vespa}, and our Illustris and EAGLE models (including dust and noise, resampled $\times \;3$).}
    \label{fig:ensemble_average_SDSS}
\end{figure}

\fig{SDSS_SFH_comparison} shows SFH predictions from\textsc{Vespa} and our Illustris and EAGLE models (trained on dust-obscured spectra, with noise resampled $\times$ 3) for four example SDSS galaxies.
We emphasise that neither our model nor the \textsc{Vespa} predicted histories represent the `true' SFH, but are shown simply to highlight the differences.
Our model SFHs are much smoother than those predicted from \textsc{Vespa}, which predicts more stochastic, bursty histories.

Our observational selection is neither mass nor volume complete, so it is not possible to make a fair evaluation of the population SFH or cosmic star formation rate density as a function of time.
However, we can plot the median SFH from each model for this selected sample to better understand the ensemble prediction, shown in \fig{ensemble_average_SDSS}.
\textsc{Vespa} predicts two large peaks in the SFR distribution at $\sim 20\,0\,\mathrm{Myr}$ and $\sim \,1\,\mathrm{Gyr}$, whereas our model predictions for Illustris and EAGLE have smoother, decreasing behaviour, peaked in the earliest bin.

\fig{ensemble_average_SDSS} also shows the median input SFH from the simulation for a sample magnitude-matched to the observations, which can be thought of as the effective `prior' on the SFH distribution.
The predicted distributions are similar, though not identical, to the training distributions, which suggests that the prior is highly informative, as expected, but does not dominate.

\begin{figure}
	\includegraphics[width=\columnwidth]{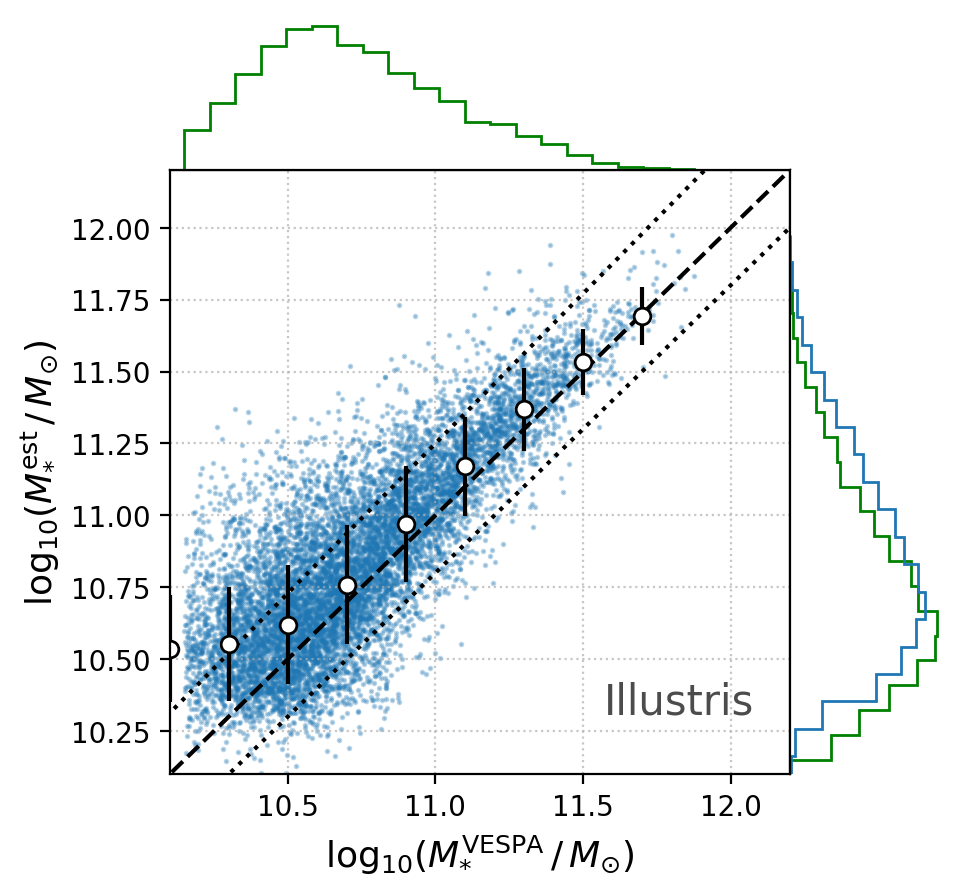}
	\includegraphics[width=\columnwidth]{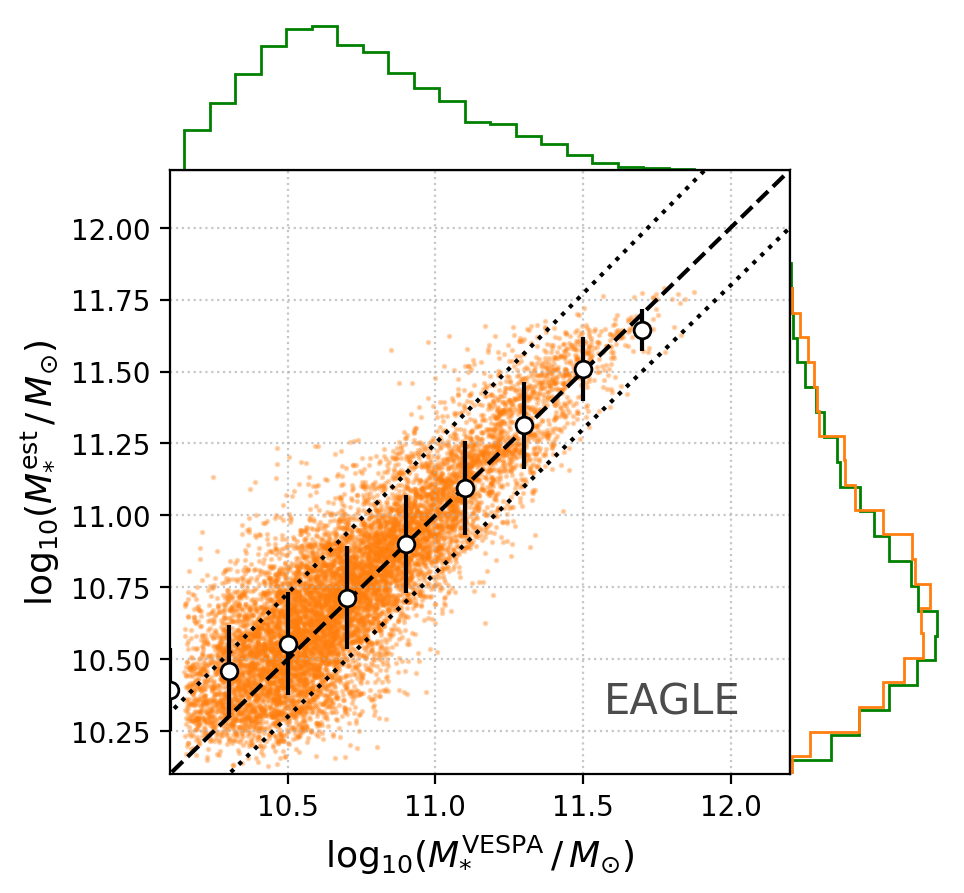}
    \caption{Estimated final stellar masses from the predicted SFH in the Illustris (top, blue) and EAGLE (bottom, orange) models, assuming an age dependent recycling fraction, compared to those published in the VESPA catalogue. The black dashed line shows the one-to-one relation, and the dotted black lines show $\pm$0.25 dex offset. The white points show the binned median and 1$\sigma$ scatter. The histograms at right show the marginal distributions of estimated stellar masses; the histogram for the VESPA distribution (green) is shown at top, and at right for comparison. The mass estimates are very similar to those obtained from \textsc{Vespa} down to $\mathrm{log_{10}}(M_{*} \,/\, M_{\odot}) \sim 10.5$, with little scatter.}
    \label{fig:SDSS_mass_vespa_comparison}
\end{figure}

We also estimate the final stellar mass of each galaxy from the SFH by assuming an age dependent recycling fraction \cite[estimated using python-FSPS;][]{dan_foreman_mackey_2014_12157}.
\fig{SDSS_mass_vespa_comparison} shows our estimates obtained from the EAGLE and Illustris models compared to the VESPA estimates.
Both models return similar stellar masses to \textsc{Vespa}, within $\sim 0.25$ dex for the majority of galaxies, and there are no mass dependent trends down to $\mathrm{log_{10}}(M_{*} \,/\, M_{\odot}) \sim 10.5$; there is a floor to the predicted masses, due to the lack of simulated galaxies with such low masses in the magnitude-selected sample.

\begin{figure}
	\includegraphics[width=\columnwidth]{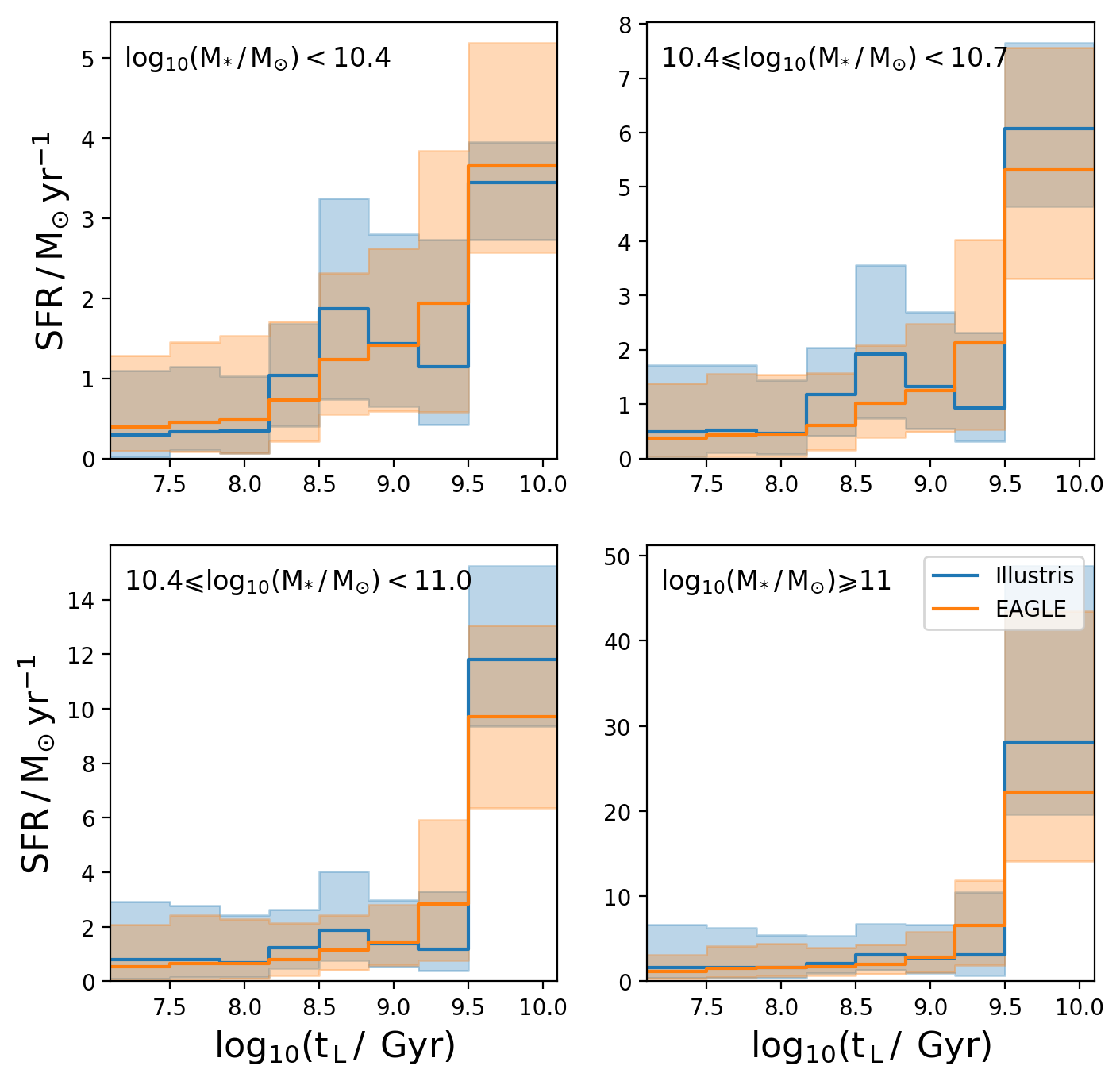}
    \caption{SDSS predictions from EAGLE and Illustris split by VESPA predicted total mass. The lines show the median, and the shaded region the 16$\mathrm{^{th}}$-84$\mathrm{^{th}}$ percentiles. EAGLE and Illustris SFH predictions for low mass galaxies are significantly different, with Illustris predicting a younger average population.}
    \label{fig:SDSS_mass_split}
\end{figure}

Finally, \fig{SDSS_mass_split} shows the median predicted SFH from each model, binned by total predicted mass (from the \textsc{Vespa} model). For EAGLE, all four bins show very similar behaviour, in both the median and the 16$\mathrm{^{th}}$-84$\mathrm{^{th}}$ percentile spread around it.
Illustris, in contrast, predicts a peak in the SFH at intermediate ages for lower mass galaxies, giving younger average stellar ages.

%% file: conc.tex
\section{Discussion}
\label{sec:disc}

We have demonstrated a new approach to estimating star formation histories using cosmological simulations, combined with detailed synthetic spectral modelling, to train a convolutional neural network. This approach is subject to different systematics and modelling assumptions compared to traditional SED fitting, which we discuss in greater detail here, as well as possible extensions in future work.

\subsection{Cosmological Simulations}
One of the obvious limitations to using cosmological simulations as a training set is that our understanding of galaxy formation is incomplete, and as such cosmological simulations are not truly representative of actual galaxies, neither individually or in ensemble, which can impact the predicted SFHs.
More realistic modelling is an obvious remedy, though this is already a fundamental aim of galaxy evolution studies.

One way of evaluating the predicted population SFH distribution is to look at the evolution of the cosmic star formation rate density (CSFRD), which in hydrodynamic simulations has been shown to be consistently in tension with observational constraints at cosmic noon ($z \sim 2$) \citep{somerville_physical_2015}.
Key distribution functions in \textsc{EAGLE} and Illustris of point-in-time properties, such as stellar mass and star formation rate, are also in tension with both observations and each other at high redshift.
Semi-analytic models are able to match these distribution functions better at a range of redshifts \citep[\textit{e.g.}][]{henriques_galaxy_2015,clay_galaxy_2015}, but do not resolve the stellar populations.

Incorrectly predicted galaxy properties also impact spectral modelling where it is physically motivated.
One physical property that has a large impact on our dust model is the central cold gas mass; both EAGLE and Illustris have been shown to underestimate this mass, to differing extents \citep{crain_eagle_2017, genel_introducing_2014}.
We find that the average star forming gas mass in Illustris is higher than in EAGLE for our selected galaxies; in EAGLE there are a significant number of galaxies with zero star-forming gas, which gives zero attenuation in our dust model ($\gamma = 0$).
This leads to higher average attenuation for Illustris galaxies, however this is cancelled out to some degree by the higher median SFR in Illustris over our mass range (see \fig{mstar_sfr_sim_comparison}), which leads to higher intrinsic luminosities.
This could possibly explain the good agreement in optical colours with observations presented in \cite{trayford_colours_2015} and \mbox{\cite{genel_introducing_2014}}, despite the differing star-forming sequence behaviour between the simulations.

\cite{trayford_colours_2015} find that, using a very similar dust model to that used in this work, EAGLE galaxies over the stellar mass range $10^{10.5} < M_{*} \,/\, M_{\odot} < 10^{10.8}$ exhibit a stronger bimodal colour distribution than that seen in observations from the GAMA survey \citep{taylor_galaxy_2015}, with higher fractions of blue galaxies.
This strong bimodal behaviour remains when the authors use an orientation dependent dust model. The colour distribution may be related to the lower passive fractions ($\sim 20\%$) seen in this mass range compared to observations \citep{schaye_eagle_2014}.
Such trends will affect predictions from the EAGLE-trained model, since its `prior' for the SFR distribution will be skewed towards more star forming objects that may not be representative of the true SFR distribution of galaxies.
Similar arguments can be made for the Illustris predictions, where the SFR distribution has a higher normalisation for intermediate masses.
We do not find that the model stellar mass estimates for SDSS galaxies show significant biases compared to VESPA, but the mass is dominated by the wide early bin.
We could of course have selected SDSS galaxies with similar stellar masses, but these may have been out-of-bounds in spectral space.
We conclude that improved physical and spectral modelling in the simulations to match the magnitude - stellar mass relations would improve our predictions.

\subsection{Spectral modelling}

The difference between synthetic spectra and observed spectra, known as the \textit{synthetic gap}, can lead to significant biases in predicted histories. More sophisticated approaches to modelling the dust could reduce this gap.
Dust models that take in to account the geometry of the gas and stars within the system show better agreement with observed colour distributions \citep{trayford_colours_2015,dave_mufasa:_2017}.
The most sophisticated approach employs 3D Monte-Carlo radiative transfer (RT), which treats absorption and anisotropic scattering by dust, as well as thermal re-emission and dust heating, in a self-consistent way. This approach has been applied to the EAGLE simulations using the \textsc{SKIRT} code, to calculate the FIR and dust properties of the galaxy population \citep{camps_far-infrared_2016, trayford_optical_2017}; they find a better match to observed local colour distributions compared to screen models.
Introducing such line of sight dependence on the attenuation is expected to reduce the correspondence between the simulated spectra and the underlying SFH, equivalent to reducing the information content of the spectra for learning our target property, the SFH.
This may lead to greater uncertainties in the derived SFHs; we will explore the effect of this in future work.

In \app{t_sne} we briefly explore the use of t-distributed Stochastic Neighbour Embedding (t-SNE) to evaluate the similarity of our synthetic spectra to the observations.
Whilst the results are good for visualisation purposes, this method is particularly sensitive to the choice of hyperparameters, such as learning rate and complexity.
\cite{masters_mapping_2015} demonstrate how self-organised maps can be used as an alternative means of addressing similarity in multi-dimensional feature spaces, whilst requiring fewer free parameters.
We plan to use this in future work as an alternative, potentially more robust way of assessing the synthetic gap.

\subsection{Machine learning approach}

We use a simple cut in stellar mass to select our training sample, which we found does not lead to overfitting of low mass galaxies despite the steepness of the GSMF (see \sec{param_correlations}).
It is unclear whether the lack of overfitting to low mass objects would extend to lower stellar masses, however the results presented here are promising.
Predictions for rare objects could also be improved by using larger volume simulations and/or `zoom' resimulations of biased regions, to increase the sampling of extreme objects, though this would negate the advantage gained from using a representative sample.

We rely on cosmological simulations for training data due to the small number of galaxies ($\sim$20) for which resolved, reasonably confident measurements of the true SFH are known.
Such objects are also mostly in the local universe, restricting any predictions to this period.
However, with ever increasing samples locally, including from from integral field unit (IFU) spectrographs \citep{bundy_overview_2015, gonzalez_delgado_spatially-resolved_2017}, it may soon be possible to train a machine on high resolution observational data in order to predict the SFH of galaxies with only unresolved data on a larger number of objects.

Our approach relies on a fixed grid of input features. Where observational data do not cover this wavelength range we currently ignore them.
An alternative to this would be to impute missing features, for example through interpolation.

\subsection{Future Extensions}

A unique aspect to our approach is that it can take advantage of the detailed modelling of complex, non-linear processes in the simulations to infer more physically motivated SFHs.
This could also be extended to other quantities self-consistently predicted in the simulations, but not directly responsible for the optical emission.
For example, halo mass could be used as a predictor, and the results compared to abundance matching approaches.
We will explore this in future work.

A powerful complement to using spectroscopic features would be to use multi-wavelength photometry, such as that available in the CANDELS fields.
However, convolution across this smaller, heterogeneous feature set would be inappropriate; using a tree based or fully connected network would lead to better performance, both computationally and predictively.
We could then compare our results to those obtained via SED fitting on photometry, using different codes such as SpeedyMC \citep{acquaviva_simultaneous_2015} and other alternatives such as Prospector \citep{leja_deriving_2017} and BEAGLE \citep{chevallard_modelling_2016}.
This will clarify how the biases and projected uncertainties of the two techniques compare, and help us make a final recommendation on improved star formation histories from multiple methods.
Tree based methods also provide information on feature importance, by equating importance with depth of features in the tree.
Deep learning based approaches necessarily obscure the relationship between the predictors and the features through the complexity of the built network, which makes it difficult to extract feature importance.

\section{Conclusions}
\label{sec:conc}

We have used convolutional neural networks (CNN) to learn the relationship between galaxies spectra and their star formation histories (SFH), using synthetic spectra generated from two cosmological hydrodynamic simulations, EAGLE and Illustris, as our training data. Our findings are as follows:

\begin{itemize}
  \item The CNN is capable of recovering the SFH of test galaxies to high accuracy (SMAPE = 10.9\%), despite the presence of dust and noise, and with no significant bias with stellar mass, SFR or stellar-metallicity.

  \item We estimate the uncertainty in our predictions from observational errors and modelling errors, and use these in combination to provide a realistic error budget on unseen data. Modelling errors dominate for both dust-obscured and intrinsic spectra.

  \item We demonstrate the good generalisation properties of the technique by applying a model trained on one simulation to simulated data from another, obtaining good accuracy (SMAPE = 14.4\% for the dust-attenuated Illustris model applied to EAGLE data) even on these unseen spectra. The model also recovers the star-forming sequence of the input data, which suggests it is not overfitting to a particular simulation.

  \item We apply our models to a magnitude matched sample of SDSS DR7 spectra and compare to the SFHs from the \textsc{Vespa} catalogue. The model predicts smoother SFHs, influenced by the `prior' distributions from the simulations, whilst recovering consistent total stellar mass predictions.

  \item When applied to our SDSS selection, the Illustris-trained model predicts younger average stellar ages for low mass galaxies ($\sim 10^{10} M_{*}\,/\, M_{\odot}$) than the EAGLE-trained model. For higher mass galaxies ($\sim 10^{10} M_{*}\,/\, M_{\odot}$) both models predict similar SFH (and hence age) distributions.

\end{itemize}

%% file: appendix.tex
\section{Correlation matrices}
\label{sec:correlation_matrices}

\begin{figure}
	\includegraphics[width=\columnwidth]{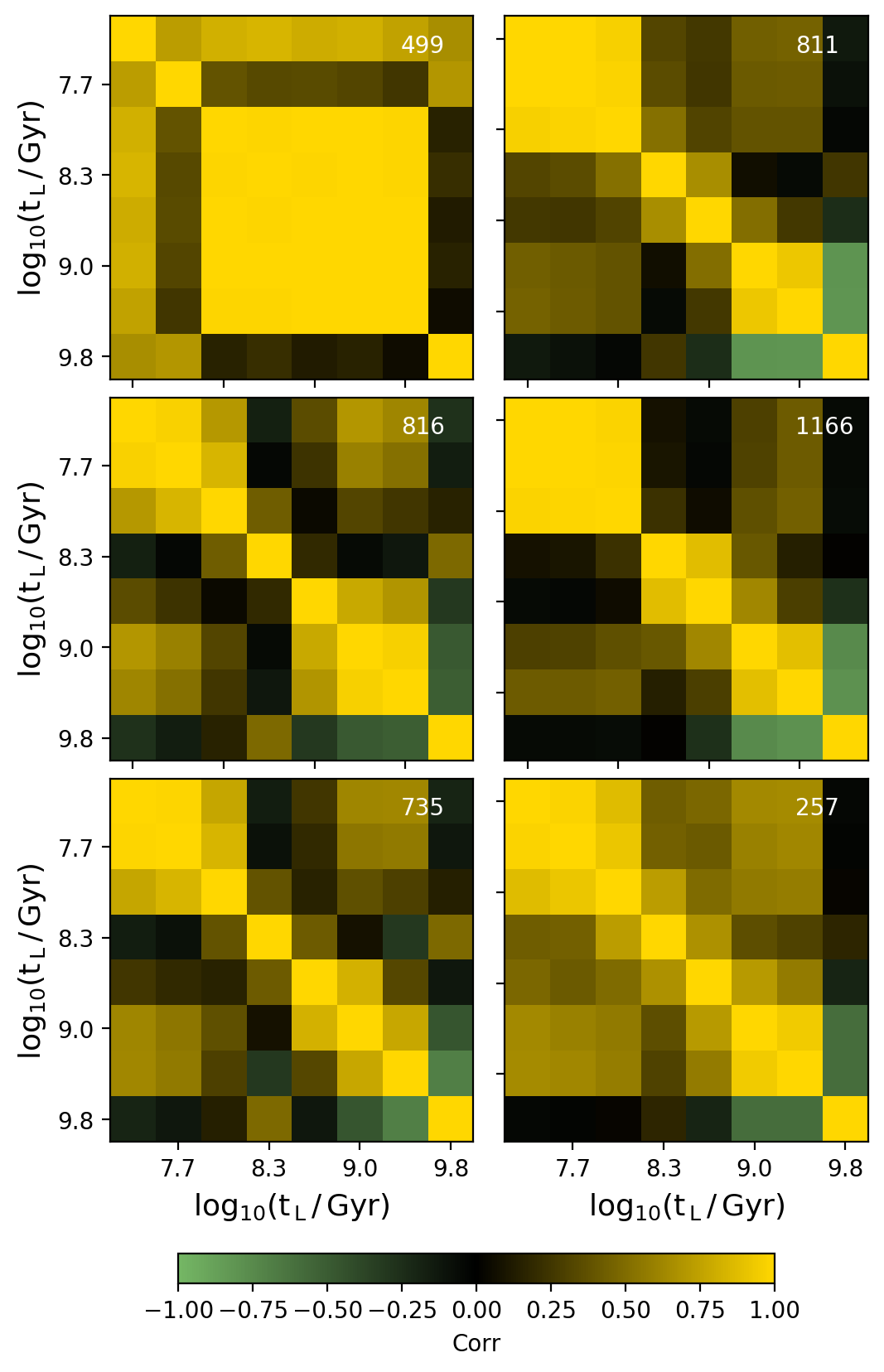}
    \caption{Correlation matrix from spectral errors, for the six galaxies shown in \fig{panel_plot} (the corresponding indices are printed in the top right corner of each panel). The colour scale varies through yellow, black and green, which show positive, neutral and negative correlation, respectively.}
    \label{fig:correlation_matrix_example}
\end{figure}

The correlation matrix can be inferred from the covariance matrix of the spectral errors, and shows the interdependence of each bin as a result of changes to the input spectra. It is given by
$$ r_{ij} = \frac{C_{ij}}{\sigma_{i}\sigma_{j}} \,\,,$$
where $r_{ij} \in [-1,1]$.
\fig{correlation_matrix_example} shows the correlation matrix for each galaxy shown in \fig{panel_plot}; the corresponding indexes are quoted in the top right of each panel.
In general, adjacent bins show the highest correlation, as expected since they are constrained by similar spectral features.
More distant bins tend to show anti-correlation, which may be due to the stellar mass constraint; where SFR increases in one bin, it is reduced in others so that the total stellar mass is reproduced.

\section{Error Tables}
\label{sec:total_errors}

\begin{table}
	\caption{Fitted parameters for the observational and modelling errors. The first two columns state the bin edges in log-lookback time. $m_2$, $m_1$ and $c$ give the second order polynomial fit parameters to the observational error. $\sigma_{\mathrm{model}}$ gives the 1$\sigma$ spread in a normal fit to the fractional residual distribution.}
	\label{tab:errors}
	\begin{tabular}{cccccc}
		\hline
			\multicolumn{2}{c}{Bins [$\mathrm{log_{10}(t_{L})}$]} & $m_{2}$ & $m_{1}$ & $c$ & $\sigma_{\mathrm{model}}$ \\
		\hline
		\hline
		 0.00 & 7.50  & 0.31 &   0.17 &  0.00 & 0.22 \\
		 7.50 & 7.83  & 0.36 &   0.10 &  0.01 & 0.19 \\
		 7.83 & 8.17  & 0.39 &  -0.02 &  0.02 & 0.17 \\
		 8.17 & 8.50  & 0.32 &  0.03  &  0.02 & 0.16 \\
		 8.50 & 8.83  & 0.39 &  0.01  &  0.03 & 0.21 \\
		 8.83 & 9.17  & 0.58 &  -0.15 &  0.06 & 0.21 \\
		 9.17 & 9.50  & 0.63 &  -0.12 &  0.09 & 0.24 \\
		 9.50 & 10.10 & 0.63 &  -0.47 &  0.16 & 0.09 \\
		 \hline
	\end{tabular}
\end{table}

In \sec{errors} we describe our method for estimating the uncertainty in the SFH predictions from \textit{observational} and \textit{modelling} errors. In \sec{spec_errors} we fit second-order polynomials to the mean observational error distribution in each bin, for dust-obscured spectra,
$$e_{\mathrm{exp}} = m_{2}x^{2} + m_{1}x + c \;\;.$$
The fit parameters are shown in \tab{errors}.
In \sec{model_uncertainty} we fit the fractional residual distribution with a normal; the $1\sigma$ spread is quoted in \tab{errors}. To obtain the $1 \sigma$ modelling error simply multiply the predicted SFR in each bin by $\sigma$. To estimate the \textit{total} error in each bin, add the observational and modelling errors in quadrature.
The distribution of fractional residuals is slightly non-symmetric, resulting in an over-estimate of the average error; we have tested the effect of this by measuring the fraction of the true SFH that lies within the errors, and found that this is the case for approximately 70\% of cases, close to the 1$\sigma$ definition.

\section{t-distributed Stochastic Neighbour Embedding}
\label{sec:t_sne}

\begin{figure*}
    \includegraphics[width=\textwidth]{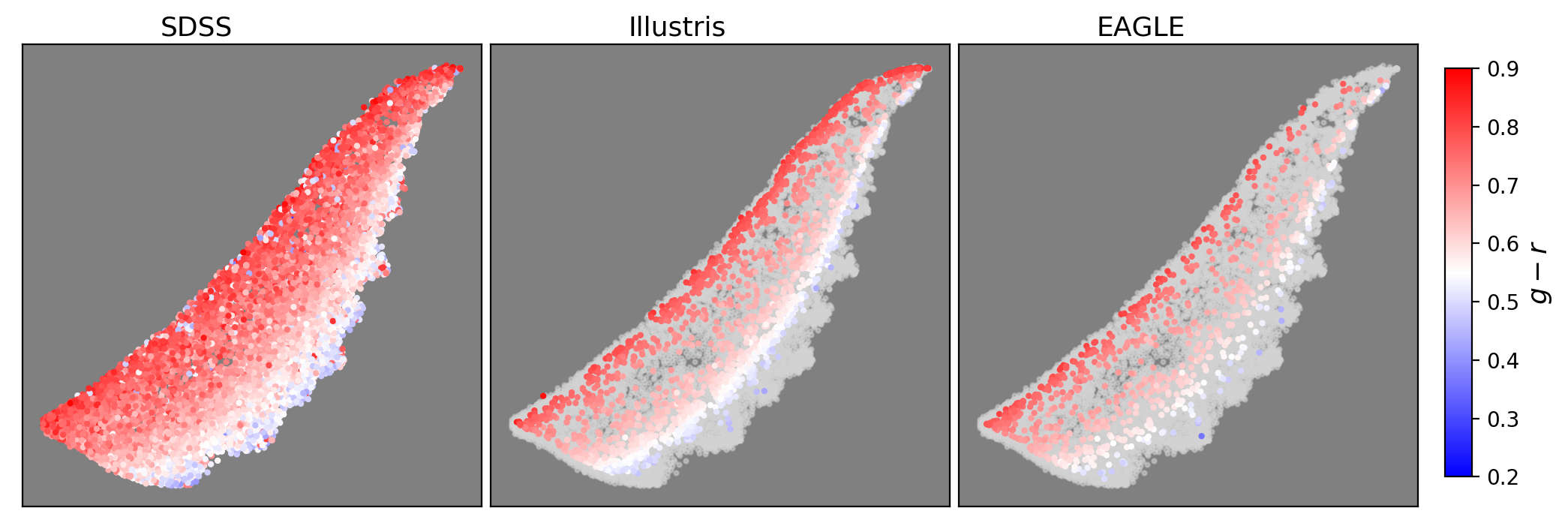}
    \caption{$t$-SNE plot applied to spectra from the SDSS selection (left panels) and the Illustris (middle panels) and EAGLE (right panels) selections. Each point represents a single galaxy spectrum. Nearby points in this 2D space have high spectral similarity. Each distribution is coloured by $g-r$ colour. The SDSS selection is shown in the background in light grey for the middle and right panels for comparison. }
    \label{fig:t_SNE}
\end{figure*}

In order to generate robust predictions using a supervised machine learning model, one needs confidence that data used to train the model are representative of the data to which it is to be applied.\footnote{One approach, proposed by \cite{cohn_multiwavelength_2019} in the context of galaxy cluster mass estimation, is to compare inferred correlations between observables in the simulations to those in actual observables.}
Synthetic spectra will always exhibit a bias compared to observational spectra, known as the \textit{synthetic gap}; where it is large it can limit the applicability of learning algorithms trained on synthetic data to observations.
To evaluate the synthetic gap we use $t$-distributed stochastic neighbour embedding ($t$-SNE) \citep{Maaten_2008}, a technique for reducing high dimensional data down to a lower number of dimensions whilst preserving the multi-dimensional distance, for visualisation purposes \citep{wattenberg2016how}.

\fig{t_SNE} shows the result of running $t$-SNE on the observationally matched sample of synthetic spectra, and the observations themselves.
The EAGLE and Illustris spectra are clustered in very similar regions of the two dimensional space, which suggests they exhibit very similar spectra.
We emphasise that t-SNE evaluates the synthetic gap across the whole of the feature space; close correspondence in this space suggests very close spectral similarity.
The observational results overlap with the simulations well, though there are certain regions, particularly at the edges of the 2D distribution, where they cluster separately from the simulation distributions, suggestive of a synthetic gap.
\fig{t_SNE} shows each distribution coloured by $g-r$ colour; where the simulations and the observational spectra do not overlap in this distribution tends to be in the extremes of the colour distribution.
This may be due to the limited volume of the simulations used for training ($\mathrm{\sim 10^{6} \, Mpc^{3}}$), which will sample fewer extreme objects, such as those in dense cluster environments.
More sophisticated approaches to spectra generation (e.g. full radiative transfer) will enhance the physical realism of the synthetic spectra, and may also reduce this synthetic gap (see \sec{disc}).